# High-harmonic generation during ultrafast melting of a Mott insulating state

Ryohei Ikeda[1,2], Yuta Murakami[3,4], Daiki Sakai[1], Tatsuya Miyamoto[1,5], Toshimitsu Ito[6], Hiroshi Okamoto[1,7]

[1]*Department of Advanced Materials Science, University of Tokyo, Chiba 277-8561, Japan*

[2] *Institute of Solid states Physics, University of Tokyo, Chiba 277-8561, Japan*

[3]*Institute for Materials Research, Tohoku University, Sendai 980-8577, Japan*

[4]*Center for Emergent Matter Science, RIKEN, Wako, Saitama 351-0198, Japan*

[5]*Department of Engineering, Nagoya Institute of Technology, Nagoya, 466-8555, Japan*

[6]*National Institute of Advanced Industrial Science and Technology, Tsukuba 305-8565, Japan*

[7]*Department of Physics, Nihon University, Tokyo 101-8308, Japan*

**Solids in an intense laser field show high-harmonic generation (HHG), which can provide information on carrier dynamics and band structures in weakly correlated systems. In strongly correlated systems, a laser field can induce a transition between the various electronic phases formed by the entanglement of charge, spin, and orbital degrees of freedom via carrier generation. The HHG accompanying this process can reflect information on the nonequilibrium electronic-state dynamics along the oscillating field—an aspect that remains unresolved to date. Here, we show that an intense mid-infrared (MIR) pulse induces melting of a Mott-insulator state in a one-dimensional cuprate, $Sr_2CuO_3$, the evolution of which leaves a trace in the HH spectra. When the electric-field amplitude exceeds 6 MV/cm, carriers are efficiently generated and each harmonic frequency decreases from odd multiples of the MIR frequency. Dynamical mean-field theory (DMFT) presents a possibility that these redshifts originate from a series of electronic-structure reconstructions in each electric-field cycle during the melting of the Mott-insulator state, which modifies the radiation phase from carrier recombination cycle-by-cycle. This phenomenon is negligible in rigid-band systems. This experimental-theoretical study suggests that HH spectroscopy has the potential to obtain information on the cycle-by-cycle dynamics of nonequilibrium transitions in correlated materials.**

High harmonic generation (HHG), a nonperturbative nonlinear optical phenomenon originating from charge dynamics under an intense light field[1,2], was initially observed in atomic and molecular gases and utilised to produce attosecond pulses[3,4] and probe attosecond-scale electron dynamics[5,6]. Subsequently, the scope of HHG research has expanded to semiconductors[7-10] and semimetals[11,12]. HHG, which is closely related to the band structure, enables band dispersion[13-15], Berry curvature[16,17], and dipole matrix element[18] analyses through HH spectral measurements in weakly correlated systems with rigid-band structures. Here, we aim to access information on dynamic electronic-state changes, rather than static band structures, through HHG, focusing on strongly correlated systems.

In strongly correlated systems, couplings between the charge, spin, and orbital degrees of freedom often result in multiple competing ordered phases[19]. Capturing the dynamics of these degrees of freedom during phase transitions remains an important challenge. To date, this challenge has been addressed with a pump–probe setup[20-25] where the pump pulse with a photon energy exceeding the band gap generates charge carriers to trigger a phase transition, and the probe pulse detects its evolution. Recently, strong sub-gap excitation with a mid-infrared (MIR) pulse, which can create carriers through nonlinear excitation processes and induce phase transitions, has become available[26,27]. Intense MIR pump pulses also cause HHG, which can reflect the dynamics of elementary excitations that emerge in strongly correlated systems[28-30]. Thus, HHG might provide information about the ultrafast electronic-state changes or nonequilibrium phase transitions in such systems under an MIR electric field[31]. This approach is different from the previous proposals of HH spectroscopy on ultrafast phase transitions in correlated materials where HHG was used as a probe or measured by an advanced sub-cycle detection technique[32,33].

To clarify the features of HHG that reflect ultrafast electronic-state changes in correlated materials, we focused on a one-dimensional (1D) Mott insulator with one orbital and one electron per site. In Mott insulators, a large on-site Coulomb repulsion $U$ prevents electron motion from the singly occupied configuration, and each electron spin is arranged antiferromagnetically (Fig. 1a: lower left). The charge carriers (elementary excitations) in Mott insulators, doublons (doubly occupied sites) and holons (non-occupied sites), are created via quantum-tunnelling processes under an intense sub-gap light field and suppress or destroy the charge-localized state, thereby inducing the melting of the Mott-insulating state[26,27,34-36]. Owing to spin–charge separation in a 1D system with a large $U$[37], the spectral weight of the optical transition beyond the Mott gap is necessarily transferred to the Drude component[21,38] in the optical conductivity (Fig. 1a: middle). Here, we selected the typical 1D Mott insulator $Sr_2CuO_3$ as the target. We have systematically studied the MIR electric-field dependence of HH spectra and carrier generation processes and found that the redshifts of the HH peaks occur when carriers are driven by the MIR electric field (Fig. 1a: lower right). Using the simulations of the HH spectra based upon DMFT, we discuss the interrelation between the Mott-insulator melting owing to carrier generation and anomalous behaviours of the HH spectra.

## Results

### Electronic structure and linear optical spectrum in $Sr_2CuO_3$

The crystal structure of $Sr_2CuO_3$ comprises $CuO_4$ units arranged along the $b$-axis sharing two diagonal oxygen atoms to form CuO chains (Fig. 1b)[39]. Although the $d$ band created by the overlap of the copper $d_{x^2-y^2}$ orbital with the oxygen $p$ orbital is half-filled (Fig. 1c), the system is a Mott insulator because of the large Coulomb repulsion $U_\mathrm{d}$ between the $d$ electrons. The O-$2p$ band is located between the upper and lower Hubbard bands consisting of Cu-$3d$ orbital, and the system functions as a charge-transfer (CT) insulator (Fig. 1d). Therefore, its electronic state should be described by the two-band Hubbard model considering O-$2p$ and Cu-$3d$ orbitals. As theoretically demonstrated, however, by mapping the O-$2p$ band to the lower Hubbard band, effectively, the electronic state of this compound can be discussed using the single-band Hubbard model[40-44]. In the optical conductivity ($\sigma$) spectrum of $Sr_2CuO_3$ obtained from the polarised reflectivity ($R$) spectrum along the $b$-axis (Fig. 2a), the sharp peak at 1.76 eV corresponds to the CT-gap energy, which is labelled the Mott-gap energy $\Delta_\mathrm{Mott}$ in this study. More strictly, the lowest excited state responsible for the peak is the one-photon allowed exciton[45], and a doublon-holon continuum exists slightly above it in energy[46]. No structures due to additional carriers are observed below $\Delta_\mathrm{Mott}$ in the $\sigma$ spectrum, indicating that the sample is a clean half-filled Mott insulator.

### Spectra of HHs by MIR excitations in $Sr_2CuO_3$

On irradiating $Sr_2CuO_3$ with MIR pulses with a photon energy of $\hbar\Omega = 0.263$ eV (frequency of $\Omega/2\pi = 63.6$ THz) and a maximum amplitude $|E_\mathrm{MIR}|$ of ~4.0–12.3 MV/cm polarised along the $b$-axis, HHs polarised along the $b$-axis were detected in a reflection configuration (Extended Data Fig. 1). The HHs polarised perpendicular to the $b$-axis were negligibly small (Supplementary Information S2). Figure 2b shows the HH spectra in which odd-order harmonics from the 3rd to 11th, 13th, and 15th orders are observed for $|E_\mathrm{MIR}| > 4.0,\ 5.0,$ and $8.5$ MV/cm, respectively. Because $Sr_2CuO_3$ has spatial inversion symmetry, only odd-order harmonics are observed[47]. The intensities of the harmonics decrease from the 3rd to 7th orders, remain almost constant from the 7th to 11th orders, and decrease again from the 13th order, similar to the spectral pattern of semiconductors based on the three-step model[2,7-9]. At $|E_\mathrm{MIR}| > 6.0$ MV/cm, the HH peak energies of the 5th to 13th order shift from odd multiples of $\hbar\Omega$, which reflects the Mott-insulator melting, as discussed later.

Figure 2c shows the $|E_\mathrm{MIR}|$ dependence of the integrated intensities of the $n^\mathrm{th}$-order harmonics, $I_{n\mathrm{HG}}$. For $|E_\mathrm{MIR}| < 4.0$ MV/cm, $I_\mathrm{3HG}$ and $I_\mathrm{5HG}$ follow the 6th and 10th power of $|E_\mathrm{MIR}|$ as $I_{n\mathrm{HG}} \propto |E_\mathrm{MIR}|^{2n}$, respectively, suggesting that they originate from multiphoton processes. At higher electric-field amplitudes, $I_\mathrm{3HG}$ is significantly enhanced, and $I_\mathrm{5HG}$ becomes slightly larger than the value suggested by the power-law dependence. When $|E_\mathrm{MIR}| > 7.0$ MV/cm, $I_\mathrm{3HG}$ and $I_\mathrm{5HG}$ tend to saturate. $I_{n\mathrm{HG}}$ for $n = 7 - 13$ do not follow the power law in the low-electric-field region, suggesting the perpetual involvement of carrier-generation processes in the HHG. Figure 2d shows the $|E_{\mathrm{MIR}}|$ dependence of $I_{n\mathrm{HG}}$ normalised by the values at $|E_{\mathrm{MIR}}| = 12.3\ \mathrm{MV/cm}$, which are qualitatively different for $n = 3$ and $n = 7 - 13$, being intermediate between them for $n = 5$. Since $3\hbar\Omega$ is smaller than the original Mott-gap energy ($\Delta_{\mathrm{Mott}}$=1.76 eV), the deviation of $I_{3\mathrm{HG}}$ from $I_{3\mathrm{HG}} \propto |E_{\mathrm{MIR}}|^6$ is attributable to doublon–holon generation and the subsequent transient currents driven by the MIR electric field, analogous to the intraband currents in semiconductors. On the other hand, the energies of $n\hbar\Omega$ for $n \geq 7$ are larger than $\Delta_{\mathrm{Mott}}$, so that they can be ascribed to the recombination of doublons and holons generated by quantum tunnelling processes and driven by the electric field, i.e., to the three-step dynamics of doublon–holon pairs. The previous theoretical studies predict a large contribution of this phenomenon towards HHG in half-filled Mott insulators[28,29]. Since the energy of $5\hbar\Omega$ is close to the absorption edge (Fig. 2a), the deviation of $I_{5\mathrm{HG}}$ from $I_{5\mathrm{HG}} \propto |E_{\mathrm{MIR}}|^{10}$ might be attributed to the three-step dynamics of doublon-holon pairs in addition to the transient currents.

**Carrier generation investigated by pump-probe reflection measurements in $Sr_2CuO_3$**

Carrier generation induced by the MIR electric field is expected to melt the Mott insulating state and modify the mechanism of HHG. To clarify this, we investigated the $|E_{\mathrm{MIR}}|$ dependence of the carrier-generation process by measuring the reflectivity changes using pump–probe reflection spectroscopy (Fig. 3a). It is known that, when carriers are doped into a 1D Mott insulator, the spectral weight of the original optical absorption across the Mott gap is transferred to the carrier-induced Drude response[21,38]. Assuming such a spectral-weight transfer, information on the carrier-generation process can be obtained from the reduction in the reflectivity at the peak corresponding to the Mott gap.

Fitting the spectra of $R$ and the imaginary part of the dielectric constant, $\varepsilon_2$ (solid lines in Figs. 3b and c, respectively) with two Lorentz oscillators representing transitions to the odd-parity exciton and doublon–holon continuum (blue and yellow broken lines in Fig. 3c, respectively) results in the purple broken lines shown in Figs. 3b,c. The theoretical study suggested that the reduction in these optical transitions is roughly proportional to the carrier density[38].

Figure 3d shows the time characteristics of the reflectivity changes, $\Delta R(t)/R$, at the reflectivity peak (1.79 eV) caused by an MIR pulse (0.264 eV) for $|E_{\mathrm{MIR}}| = 2.0 - 12.0\ \mathrm{MV/cm}$ in 1.0 MV/cm steps. For $|E_{\mathrm{MIR}}| \leq 4.0\ \mathrm{MV/cm}$, $\Delta R(t)/R$ shows a pulsed response centred at $t = 0\ \mathrm{ps}$. The $-\Delta R(0\ \mathrm{ps})/R$, plotted as circles in Fig. 3e, is proportional to $|E_{\mathrm{MIR}}|^2$ at $|E_{\mathrm{MIR}}| \leq 4.0\ \mathrm{MV/cm}$. Therefore, the observed signals are attributable to coherent 3rd order nonlinear responses associated with the excitonic Floquet state[48]. Because the Floquet state is formed only under an MIR electric field, it is observed as a pulsed response[48-50] (Supplementary Information S1 and S3).

In contrast, for $|E_{\mathrm{MIR}}| > 4.0\ \mathrm{MV/cm}$, a response with a finite decay time appears for $-\Delta R(t)/R$, which is attributable to carrier generation. On increasing $|E_{\mathrm{MIR}}|$ beyond 4.0 MV/cm, $-\Delta R(0\ \mathrm{ps})/R$ tends to saturate because the coherent response is disturbed by carrier generation. Since the coherent response completely disappears at $t = 0.25$ ps, $-\Delta R(0.25\text{ ps})/R$ should be used to investigate the carrier-generation phenomena. As seen in Fig. 3e (triangles), $-\Delta R(0.25\text{ ps})/R$ shows a threshold behaviour, suggesting that doublon-holon pairs are created via quantum-tunnelling processes induced by the MIR electric field. As shown by the purple solid line in Fig. 3e, the $|E_{\text{MIR}}|$ dependence of $-\Delta R(0.25\text{ ps})/R$ is almost reproduced by the following equation of the probability of doublon-holon creation, $\Gamma$, in 1D Mott insulators[34]:

$$\Gamma \propto \exp(-\pi E_{\text{th}}/|E_{\text{MIR}}|),\ \ E_{\text{th}} = \frac{\Delta_{\text{Mott}}}{2e\xi}\ . \tag{1}$$

Here, $E_{\text{th}}$, $\xi$, and $e$ are the threshold electric field, doublon–holon correlation length, and elementary charge, respectively. The used parameter values, $E_{\text{th}} = 13.4$ MV/cm and $\Delta_{\text{Mott}} = 1.76$ eV yield $\xi$~6.57 Å (~1.68 sites[41]). The magnitude of $-\Delta R(0.25\text{ ps})/R$ reaches 0.35 at $|E_{\text{MIR}}| \simeq 12.0$ MV/cm, suggesting that the Mott-insulating state is melted by carrier generation.

Estimating the $|E_{\text{MIR}}|$ dependence of the carrier density is valuable for discussing the properties of HHG. Here, we attempted a semi-quantitative estimation based on the value of $-\Delta R(0.25\text{ ps})/R$. First, the change in $\sigma$ at 0.25 ps should be estimated from $-\Delta R(0.25\text{ ps})/R$. The sum of doublon and holon density per site, $n_{\text{c}}$, can be approximated as $-\Delta I/2I$, where $I$ represents the sum of the strengths of two Lorentz oscillators (LO1, 2) obtained to reproduce the original $R$ and $\varepsilon_2$ spectra, and $\Delta I$ is the MIR electric-field-induced change in $I$. The green broken line in Fig. 3b represents the $R$ spectrum calculation when $-\Delta I/I = 0.3$. By back-calculating $-\Delta I/2I$ from $-\Delta R(0.25\text{ ps})/R$, we obtained the $|E_{\text{MIR}}|$ dependence of $n_{\text{c}}$ as squares (Fig. 3e), which exhibit the same tendency as $-\Delta R(0.25\text{ ps})/R$. Analysis suggested that $n_{\text{c}}$ reaches 0.25 at $|E_{\text{MIR}}| = 12.0$ MV/cm. Further details are provided in Supplementary Information S4. It should be noted that our estimation assumes that the spectral weight of the optical absorption across the Mott gap is entirely transferred to the carrier-induced Drude response. Therefore, the estimated absolute value of $n_{\text{c}}$ may be subject to some uncertainty. Nevertheless, the threshold-like onset of carrier generation with increasing $|E_{\text{MIR}}|$ is expected to be robust and closely correlates with the observed HH shifts.

The HH spectra are plotted against $\omega/\Omega$ for the 3rd to 13th order harmonics at $|E_{\text{MIR}}| = 6.0 - 12.0$ MV/cm in Figs. 4a–f and at $|E_{\text{MIR}}| = 4.0 - 6.0$ MV/cm in Extended Data Figs. 3a–f. We fitted each spectrum with a Gaussian profile and determined the frequencies of the $n^{\text{th}}$ order peak structures, $\omega_{\text{peak,n}}$. Figure 4h shows the $|E_{\text{MIR}}|$ dependence of the peak shifts, $(\omega_{\text{peak},n} - n\hbar\Omega)$. At $|E_{\text{MIR}}| \leq 6.0$ MV/cm, each peak energy almost coincides with $n\hbar\Omega$. At $|E_{\text{MIR}}| > 6.0$ MV/cm, the energies of the 3rd order harmonics are equal to $3\hbar\Omega$, while those of the 5th and higher order harmonics show redshifts with increasing $|E_{\text{MIR}}|$, saturating around $|E_{\text{MIR}}| = 12.0$ MV/cm. The redshift magnitude increases with $n$, reaching ~50 meV at the 11th harmonic. The $|E_{\text{MIR}}|$ dependence of these shifts follows both $\Delta R$(0.25 ps)/$R$ (Fig. 3e) and the estimated carrier density $n_c$ (Fig. 4g), suggesting that

the redshifts may originate from the successive electronic-state changes in the Mott-insulator melting. The frequency shifts that could arise from nonlinear propagation effects[51] associated with photocarrier-induced changes in the HH penetration depth and the refractive index can be confirmed by simulations to be much smaller than the experimentally observed shifts (Supplementary Information S5). Furthermore, for GaAs, a conventional semiconductor with a bandgap (1.42 eV) comparable to that of $Sr_2CuO_3$ (1.76 eV), no spectral shifts of HHs were reported in reflection-geometry measurements of nonperturbative HHG using 0.34-eV MIR pulses with an electric-field amplitude of 10 MV/cm.[52]

**Simulations of HH spectra based upon DMFT**

To investigate the relationship between the electronic-state changes and HH shifts, we simulated the dynamics of the half-filled Hubbard model in the Mott regime (see Methods). The previous theoretical study has shown the consistency of HHG properties calculated for the single-band Hubbard model and $dp$ model[53]. The main limitation of employing the former to analyze HHG in CT insulators lies in its inability to accurately capture HHG processes at photon energies far above the gap. However, the HHG regime of interest in the present study is restricted to energies up to only twice the gap, where an effective Hubbard-model description remains valid.

The time evolution was computed using time-dependent dynamical mean-field theory (tdDMFT)[54], which describes electronic-structure changes with carrier relaxation and scattering processes in strongly correlated systems, providing the time evolutions of the doublon and holon energy levels, single-particle spectra, $n_{\mathrm{c}}$, and antiferromagnetic spin-spin correlation $m_{\mathrm{AFM}}$. The following parameters are used to fit $Sr_2CuO_3$: $U = 2.6$ eV, bandwidth $W = 2.08$ eV, and bond length $b = 3.9$ Å. The temperature $T$ is set at 260 K, where $\Delta_{\mathrm{Mott}}$ is ~1.5 eV and antiferromagnetic order develops, consistent with the electronic state in $Sr_2CuO_3$. This Mott-insulator state is excited with a Gaussian electric-field pulse centred at $t = 0$ with the frequency $\Omega/2\pi = 63$ THz ($\hbar\Omega = 0.26$ eV) and temporal width $\sigma_0 = 63.3$ fs (Figs. 5a–c). The HH spectra calculated by tdDMFT are shown in Fig. S7. The deviations of the HH peak energies $\hbar\omega_{\mathrm{peak},n}$ from $n\hbar\Omega$ extracted from Fig. S7 and the calculated $n_c$ values are plotted as functions of $|E_{\mathrm{MIR}}|$ in Fig. 4j and i, respectively. These results capture the characteristic redshift associated with a sharp increase in $n_c$.

To further pursue the physical origin of this behaviour, we next analysed the evolutions of $n_{\mathrm{c}}$, $m_{\mathrm{AFM}}$, and single-particle spectra (Figs. 5d–f) (Supplementary Information S6). At $|E_{\mathrm{MIR}}| = 6.7\ \mathrm{MV/cm}$, they remain almost unchanged over time. At $|E_{\mathrm{MIR}}| = 12.0$ and 16.0 MV/cm, $n_{\mathrm{c}}$ increases and $m_{\mathrm{AFM}}$ decreases to nearly zero during the pulse. This disappearance of $m_{\mathrm{AFM}}$ suggests that the system is changed into a paramagnetic metal[19]. Correspondingly, the single-particle spectrum varies cycle-by-cycle, the original Mott gap shrinks, and the spectral weight is transferred from the lower to the upper Hubbard band with time. The population in the upper Hubbard band corresponds to the Drude weight in $\sigma(\omega)$. The observed evolution of the single-particle spectrum indicates transient changes in the dynamics of doublons and holons, which affect the radiation processes at each half cycle, eventually causing HH peak shifts (Fig. 4j).

It is noteworthy that HH peak shifts are influenced by both the intensity and phase of successive radiation. Figures 5g-i show the time dependence of radiation intensities, whose profiles beyond the original gap can be associated with the three-step dynamics of doublons and holons[28,29]. At $|E_{\mathrm{MIR}}|$ =6.7 MV/cm, the radiation is nearly periodic and strongest near the pulse centre (Fig. 5g). With stronger fields, the periodicity is violated, and the radiation weight shifts earlier (Figs. 5h,i). At first glance, the suppression of radiation around $t = 0$ appears counterintuitive in view of the shrinking gap, likely owing to the reduction in the coherence of doublon–holon dynamics as carrier generation and Mott-insulator melting progress.

After identifying the relevant time window, we evaluated the radiation phase to examine the conditions for constructive and destructive interference between successive radiations at different cycles. The radiation phase around a given time $t = t_{\mathrm{p}}$ is defined as $\mathrm{Arg}[\dot{J}(\omega,\ t_{\mathrm{p}})]$, where $\dot{J}(\omega, t_{\mathrm{p}})$ corresponds to the current derivative within the interval $t \in (t_{\mathrm{p}} - T_{\mathrm{hp}}/2,\ t_{\mathrm{p}} + T_{\mathrm{hp}}/2)$, and $\dot{J}(\omega) = \sum_l \dot{J}(\omega,\ t_{\mathrm{p}} + lT_{\mathrm{hp}})$, where $T_{\mathrm{hp}} = \pi/\Omega$ and $l$ is an integer indicating the time domain of each window. In Figs. 5j–l, we plot the difference of $\mathrm{Arg}[\dot{J}(\omega, t)]$ from a reference time $t_i = -73.2$ fs; $\Delta\mathrm{Arg}[\dot{J}(\omega, t)] \equiv \mathrm{Arg}[\dot{J}(\omega, t)] - \mathrm{Arg}[\dot{J}(\omega, t_{\mathrm{i}})]$ at $\omega = n\Omega$, and $\omega = \omega_{\mathrm{peak},n}$ $(n = 3$ and $9)$ for each half cycle. The results for $n = 3 - 11$ are shown in Fig. S10. At $|E_{\mathrm{MIR}}| = 6.7$ MV/cm, the phases for $\omega = n\Omega$ and $\omega = \omega_{\mathrm{peak},n}$ are almost equal, and they vary only slightly from cycle to cycle. At $|E_{\mathrm{MIR}}| = 12.0$ and $16.0$ MV/cm, similar behaviour is observed for $n = 3$, which corresponds to in-gap radiation. In contrast, for $n \geq 7$, a pronounced difference arises between $\omega = n\Omega$ and $\omega = \omega_{\mathrm{peak},n}$, which is attributable to the progress of Mott-insulator melting. The radiations at $\omega = \omega_{\mathrm{peak},n}$ exhibit smaller phase variations when their intensities are strong; they constructively accumulate compared with those at $\omega = n\Omega$. Notably, at $|E_{\mathrm{MIR}}| = 16.0$ MV/cm, the radiation phase at $\omega = n\Omega$ becomes nearly flat for $t > 0$; therefore, a suppression in HH peak shifts is expected. However, the shift occurs, because the radiation from the latter half of the MIR pulse, where the carrier density is increased, is suppressed due to the decoherence effect as mentioned above. In other words, the HH peak shifts originate from the melting process of the Mott insulating state and occur without closing of the Mott gap.

The distinct behaviour of the HH signals between $n = 3$ and $n \geq 7$ (Figs. 4j, S7, and S9) can be attributed to their distinct physical origins; the HHG at $n = 3$ originates from non-resonant processes within the gap in the perturbative regime and from the intraband current when sufficient charge carriers are generated, whereas that at $n \geq 7$ arises from the doublon–holon three-step mechanism. The HHG at $n = 5$ represents an intermediate case. It is reasonable that as the Mott gap shrinks the radiations from two types of mechanisms at $n = 3$ are both enhanced, while their phase remains unaffected.

**Discussion**

As detailed in the previous section, the $|E_{\mathrm{MIR}}|$ dependence of the HH shifts exhibits the following three characteristics:

1. The 3rd-order harmonic peak remains unchanged even when $|E_{\mathrm{MIR}}|$ increases up to 12 MV/cm.
2. For the $n$th-order harmonics with $n \geq 5$, threshold-like redshifts of the harmonic peaks emerge with increasing $|E_{\mathrm{MIR}}|$ above 6 MV/cm, resembling the $|E_{\mathrm{MIR}}|$ dependence of photocarrier generation efficiency deduced from the reflectivity changes (Figs. 4g,h), and then tend to saturate at higher $|E_{\mathrm{MIR}}|$.
3. The 5th-order harmonic shift is smaller than the 7th-order and higher harmonic shifts, exhibiting the intermediate nature between the 3rd and $n$th-order harmonics for $n \geq 7$.

All three features are semi-quantitatively reproduced by the DMFT simulation (Fig. 4i and j). This agreement suggests that the characteristic HH shifts in $Sr_2CuO_3$ can be attributed to the successive electronic-state changes during the melting of the Mott insulating state. Analyses of semiconductor-like models with and without the progressive band renormalization also support this picture (Supplemental Information S7).

In short, the HH redshifts can be related to the reduction of the Mott gap associated with the melting of the Mott insulating state. Although a quantitative extraction of the gap reduction requires detailed forward modelling, its approximate magnitude can be estimated in the following way. If the system is in a time-periodic state, the radiation phase at a frequency $\omega_r$ changes by $\pi - \omega_{\mathrm{r}} T_{\mathrm{hp}}$ for each half cycle, which results in constructive interference at $\omega_{\mathrm{r}} = n\Omega$ ($n$: odd number). Next, we consider a simplified situation where the modification in doublon–holon dispersion can be approximated as a uniform level shift by $\Delta\epsilon$ per $T_{\mathrm{hp}}$. Within the three-step model[29], the additional phase shift for the radiation from doublon–holon recombination is approximately $(\Delta\epsilon/\hbar)(t_{\mathrm{r}} - t_{\mathrm{c}})$, where $t_{\mathrm{r}}(t_{\mathrm{c}})$ is the recombination (creation) time of the doublon–holon pair. For constructive interference, the additional phase must be compensated with the phase shift due to the frequency deviation from the integer harmonic, $\Delta\omega_{\mathrm{r}} = \omega_{\mathrm{r}} - n\Omega$. Thus, the gap reduction ($\Delta\epsilon < 0$) associated with the Mott-insulator melting leads to HH redshifts ($\Delta\omega_{\mathrm{r}} < 0$). The observed $\Delta\omega_r$ is consistent in order of magnitude with $|\Delta\epsilon|/\hbar$, since typically $t_{\mathrm{r}} - t_{\mathrm{c}} \sim T_{\mathrm{hp}}$.

Control of the spectrum of HHs generated by photoexcitation in solids has primarily been achieved by varying the carrier envelope phase (CEP) of the excitation light[8,55-57]. For example, the combination of long scattering time, Dirac dispersion, and short-pulse excitation enables the generation of tunable non-integer harmonics from the surface state of the topological insulator $Bi_2Te_3$[57]. In these phenomena, *the shape of the excitation pulse* governs electron dynamics. On the other hand, the HH-shift mechanism observed in $Sr_2CuO_3$ differs from that reported previously, likely owing to the cycle-by-cycle modification of charge dynamics during the progress of nonequilibrium electronic-state changes.

Unlike CEP-dependent HHG phenomena reported previously, the present mechanism of HH redshift is insensitive to CEP (Supplementary Information S6) and does not require long scattering times.

In this study, we have revealed that HHs generated in the 1D Mott insulator $Sr_2CuO_3$ under MIR excitation exhibit redshifts arising from the melting of the Mott insulating state, whose electric-field dependence closely follows that of carrier generation deduced from transient reflectivity measurements. Furthermore, theoretical simulations showed that carrier generation and the periodic driving of the generated carriers by the MIR electric field progressively destabilize the Mott-insulating state and that the resulting reduction of the Mott gap leaves a signature in the form of HH redshifts. Our study provides a benchmark demonstration that HH spectroscopy may offer a complementary approach to conventional pump–probe spectroscopies for accessing sub-cycle electronic-state evolution in correlated materials. What fingerprints would appear in HH spectra during the light-induced melting of charge, orbital, or magnetic order, or during light-induced superconductivity? The combined experimental–theoretical approach demonstrated here provides a promising route toward addressing these questions and advancing ultrafast photonic functionalities in correlated materials.

**Acknowledgements**

We thank Mr. Z. Zhong, Mr. Y. Suzuki, and Ms. M. Xu for their assistance in constructing the optical setups and Dr. S. Kito for his support in X-ray diffraction measurements to determine the crystallographic orientation of single crystals. We also thank Prof. M. Eckstein for enlightening discussions. This work was partly supported by a Grant-in-Aid for Scientific Research from the Japan Society for the Promotion of Science (JSPS) (Project Numbers: JP21H04988, JP21H05017, JP24H00191, JP25K07235, JP25K17961, JP26K00648, JP26H01281, JP26K00646), CREST (JPMJCR1661), Japan Science and Technology Agency, Iketani Science and Technology Foundation (0371180-A), and Toyota Riken Scholar. D.S. was supported by Support for Pioneering Research Initiated by Next Generation of Japan Science and Technology Agency (JST SPRING) (Grant Number: JPMJSP2108).

**Author contributions**

R.I., D.S., T.M., and H.O. built the experimental setups of high harmonic spectroscopy and pump probe reflection spectroscopy. R.I. and D.S. performed the measurements. T.I. prepared the sample. Y.M. performed the DMFT calculations. H.O. coordinated the study. The manuscript was written by R.I., Y.M., and H.O. with inputs from the other authors.

**Competing interests**

The authors declare no competing interests.

## Additional information

**Supplementary information** is available for this paper at https://doi.org/.

**Correspondence and requests for materials** should be addressed to H.O. or Y.M.

**Reprints and permissions information** is available at.

## Methods

### Sample preparations

Single crystals of $Sr_2CuO_3$ were grown via the laser-diode-heated floating-zone method[58] using the traveling solvent technique[39]. The initial composition of the solvent with SrO:CuO = 40 mol%:60 mol% was determined from the phase diagram of the SrO–CuO system[59]. A mixture of argon (60 vol%) and oxygen (40 vol%) flowed through the growth chamber at a rate of 100 $cm^3$/min. The growth rate was 1 mm/h.

### Polarised reflectivity spectrum measurements

The polarized reflectivity ($R$) spectrum of $Sr_2CuO_3$ along the $b$ axis was measured using a Fourier transform infrared spectrometer (0.08–1.2 eV) and a spectrometer with a grating monochromator (0.46–6.2 eV), both of which were equipped with an optical microscope. The optical conductivity ($\sigma$) spectra and the imaginary part of dielectric constant ($\varepsilon_2$) spectra were calculated from the $R$ spectrum using the Kramers–Kronig transformation. All the optical measurements were performed on a clean surface along the *bc*-plane obtained by cleaving the single crystal. To prevent sample degradation, the single crystal was put in a vacuum during the measurements.

### HH spectrum measurements

Schematics of the HH spectrum measurements are shown in Extended Data Fig. 1. A Ti:sapphire regenerative amplifier (RA) with a photon energy of 1.55 eV, temporal width of 35 fs, output fluence of 7 mJ, and repetition rate of 1 kHz was used as the laser source. The RA output was used as input into an optical parametric amplifier (OPA1) to generate signal and idler lights through an optical parametric process; these outputs were input into a 1 mm-thick second-order nonlinear optical crystal, z-cut GaSe, to generate an MIR pulse with a photon energy of 0.263 eV via a differential-frequency generation process. The signal and idler lights used to generate the MIR pulse were cut using long-pass filters (LPF1 and LPF2) to ensure that they were not introduced into the sample. The maximum electric-field amplitude of the MIR pulse was controlled using two MIR wire-grid polarisers (WG1 and WG2), and a third polariser (WG3) was used to ensure good light polarisation along the *b*-axis of $Sr_2CuO_3$. The temporal and spectral widths of the MIR pulses were ~220 fs and 22 meV, respectively (Supplementary Information S1).

To measure the HH spectra in the normal-reflection configuration, a soda-lime glass substrate deposited with indium tin oxide (ITO) was used as a beam splitter (BS) for MIR and HH light. A broadband wire-grid polariser (WG4), applicable in the near-IR (NIR) to visible (VIS) region, was used to resolve the polarisation of the HH light. Two fiber spectrometers, an NIR spectrometer (Lambda Vision) and a VIS spectrometer (QE-Pro, Ocean optics), were used to measure the HH spectra in the 0.70–1.5 eV and 1.4–4.0 eV regions, respectively. The measured HH spectra were modified by the optical properties of the optics shown in Extended Data Fig. 1 and the detection sensitivities of the two spectrometers, both of which depend on the wavelength of the HH light. Therefore, the HH spectra should be corrected. By placing a standard light source, a halogen tungsten lamp (DH-3 PLUS, Ocean Optics), at the sample position and comparing the obtained spectrum with the spectrum of the standard light source, we evaluated the correction factor, which was used to correct all the HH spectra.

**MIR pump-visible reflectivity probe measurements**

Schematics of the MIR pump-visible reflectivity probe measurements are shown in Extended Data Fig. 2. The laser source and generation method for the MIR pump pulses were identical to those used for the HH spectrum measurements. The photon energy was 0.264 eV. A part of the RA output was input into another OPA2, from which the probe pulse (1.79 eV) was generated. The temporal and spectral widths of the visible pulses were 55 fs and 55 meV, respectively, while those of the MIR pulses were ~170 fs and 24 meV, respectively (Supplementary Information S1). The delay time of the probe pulse relative to the pump pulse was varied using a mechanical stage installed in the probe beam path. The MIR-pump and VIS-probe pulses were polarised parallel to the *b*-axis of $Sr_2CuO_3$. To reduce the effects of intensity fluctuations of the probe pulses, a part of each probe pulse was extracted by a BS, and its intensity was measured as a reference with photodetector 2 (PD2). The intensity of the probe pulse reflected from the sample was measured as a signal with photodetector 1 (PD1) and normalised by the reference using a boxcar integrator.

**DMFT calculations**

We analysed the time evolution of the single-band Hubbard model to understand the origin of the redshifts of HH peaks. Under an electric field, the Hubbard model can be expressed as follows:

$$\hat{H}(t) = -t_{\text{hop}} \sum_{\langle i,j \rangle} e^{i\phi_{ij}(t)} \hat{c}_{i\sigma}^{\dagger} \hat{c}_{j\sigma} + U \sum_i \hat{n}_{i\uparrow} \hat{n}_{i\downarrow},$$

where $\hat{c}_{i\sigma}^{\dagger}$ is the creation operator for an electron with spin $\sigma$ at site $i$, $\langle i,j \rangle$ indicates a pair of neighbouring sites, $\hat{n}_{i\sigma} = \hat{c}_{i\sigma}^{\dagger}\hat{c}_{i\sigma}$, $t_{\text{hop}}$ is the hopping parameter, and $U$ is the onsite Coulomb repulsion. The effect of electric fields is included via a Peierls phase, $\Phi_{ij}(t) = -e\boldsymbol{A}(t) \cdot \boldsymbol{r}_{ij}$. Here, $\boldsymbol{A}(t)$ is the vector potential, which is related to the electric field $\boldsymbol{E}(t)$ as $\boldsymbol{E}(t) = -\partial_t \boldsymbol{A}(t)$ and $\boldsymbol{r}_{ij}$ is the position operator from site $j$ to site $i$. Here, $U$ (2.6 eV), the bandwidth ($W =$ 2.08 eV), and the bond length ($b = 3.9$ Å) were set to be consistent with $Sr_2CuO_3$, unless otherwise mentioned. The temperature was set as $T = 260$ K.

We solved this model using a nonequilibrium extension of the DMFT[53], a powerful numerical method for studying strongly correlated systems both in and out of equilibrium. DMFT is based on Green's function formalism, where the nonequilibrium Green's function $G_i(t,t') = -i\langle T_C \hat{c}_i(t)\hat{c}_i^\dagger(t')\rangle$ is defined on the L-shaped contour, *C*. In DMFT, the original lattice model is mapped to an effective impurity model to fully incorporate dynamic local correlations. This method is suitable for systems in the thermodynamic limit and can explain the Mott physics, reconstruction of electronic structures, and relaxation or scattering processes. In this study, we used the non-crossing approximation as the impurity solver within the DMFT self-consistency loop; the code was based on the open-source library Nessi[60].

For simulations, we used the Hubbard model on a Bethe lattice to reduce computational cost and ensure systematic analysis. This setup corresponds to the application of an electric field along the body diagonal of a general hypercubic lattice. Previous studies have shown that DMFT dynamics are generally insensitive to lattice geometry, and similar behaviour is expected in other lattices. We set the pump pulse to a Gaussian pulse as follows:

$$A(t) = \frac{E_0}{\Omega} F_{\mathrm{Gauss}}(t - t_0, \sigma_0) \sin[\Omega(t - t_0) + \phi_{\mathrm{CEP}}],$$

where $F_{\mathrm{Gauss}}(t, \sigma_0) = \exp[-t^2/2\sigma_0{}^2]$. We set $\hbar\Omega = 0.26$ eV, $t_0 = 0$ fs, $\sigma_0 = 63.3$ fs, and $\phi_{\mathrm{CEP}} = 0$, and simulated from $t_{\min} = -253.2$ fs up to $t_{\max} = 253.2$ fs, unless otherwise mentioned. Within tdDMFT, we evaluated various physical observables to characterise the dynamics of the system. The HHG spectrum $I_{\mathrm{HHG}}(\omega)$ is computed from the current $J(t)$ using $I_{\mathrm{HHG}}(\omega) = |\dot{J}(\omega)|^2 = |\omega J(\omega)|^2$. Here, $\dot{J}(t) \equiv \partial_t J(t)$ is the time derivative of the current, and $J(\omega)$ and $\dot{J}(\omega)$ are the Fourier transform of $J(t)$ and $\dot{J}(t)$. In practice, to reduce noise in the Fourier transform due to the finite time range of the simulation, we apply a Gaussian window $F_{\mathrm{Gauss}}(t - t_0, \sigma')$. Here, $\sigma'$ is chosen to be wider than $\sigma_0$ but shorter than $t_{\max} - t_0$. In practice, we set $\sigma' = 101.3$ fs to satisfy this condition. Notably, the results were found to be insensitive to the choice of $\sigma'$.

To resolve the radiation behaviours at each half cycle within the pump pulse, we applied a windowed Fourier transform $\dot{J}(\omega, t_{\mathrm{p}}) = \int dt e^{i\omega t} F_{\mathrm{window}}(t - t_{\mathrm{p}}) \dot{J}(t)$, which provides an emission profile around $t_{\mathrm{p}}$, where $F_{\mathrm{window}}(t)$ is a window function centred at $t = 0$. For the time-dependence of the intensity spectrum in Figs. 5g–i, we set $F_{\mathrm{window}}(t) = F_{\mathrm{Gauss}}(t - t_{\mathrm{p}}, \sigma_{\mathrm{p}})$ with $\sigma_p = 1.5$ fs and evaluated $I_{\mathrm{HHG}}(\omega, t_{\mathrm{p}}) \equiv |\dot{J}(\omega, t_{\mathrm{p}})|^2$. To extract the phase information of the emission cycle-by-cycle in Figs. 5j–l, we set $F_{\mathrm{window}}(t)$ as $F_{\mathrm{box}}((t - t_{\mathrm{p}})/T_{\mathrm{hp}}, 0.2)$, which covers a half cycle of the pulse $t \in (t_\mathrm{p} - T_\mathrm{hp}/2, t_\mathrm{p} + T_\mathrm{hp}/2)$. A detailed description of $F_\mathrm{box}$ is provided in Supplementary Information S6.

The single-particle spectrum represents the electronic states that are accessible by adding or removing one electron to or from the system, and its occupation can be measured by photoemission experiments. In practice, they can be expressed as follows:

$$A(t_\mathrm{p}, \omega) = -\frac{1}{\pi}\mathrm{Im}\left[\int dtdt' e^{i\omega(t-t')} S(t)S(t')G^R(t + t_\mathrm{p}, t' + t_\mathrm{p})\right],$$

$$N(t_\mathrm{p}, \omega) = -\frac{i}{2\pi}\int dtdt' e^{i\omega(t-t')} S(t)S(t')G^<(t + t_\mathrm{p}, t' + t_\mathrm{p}),$$

where $G^\mathrm{R}$ and $G^<$ are the retarded and lesser parts of the (on-site) Green's function, respectively, and $S(t)$ is a Gaussian window function representing the probe pulse in the photoemission experiment. In our analysis, we set the width of $S(t)$ to 5 fs to capture the dynamics within each half cycle. These functions are local and averaged over the sites. Notably, the occupation in the upper Hubbard band corresponds to doublon formation. Thus, the doublon density was evaluated by integrating the occupation over the upper Hubbard band, as follows:

$$n_\mathrm{d}(t_\mathrm{p}) \equiv \frac{\int_0^\infty d\omega N(t_\mathrm{p}, \omega)}{\int_{-\infty}^\infty d\omega A(t_\mathrm{p}, \omega)}.$$

Because the number of holons equals the number of doublons in this system, the total carrier density per site, $n_\mathrm{c}$, is $2n_\mathrm{d}(t)$. The antiferromagnetic spin–spin correlation $m_\mathrm{AFM}$ is defined as $n_{\uparrow,\mathrm{A}} - n_{\uparrow,\mathrm{B}}$, where A and B represent two sublattices in the bipartite lattice.

**Summary on time resolutions of the DMFT simulation**

In Fig. 5, which displays the DMFT results, multiple datasets with differing time resolutions are included. Those time resolutions are summarized as follows. The current $J(t)$ required for the HHG analysis and for evaluating the antiferromagnetic order parameter can be calculated with the intrinsic accuracy of the time-evolution step, which is approximately 0.03 fs. In contrast, the time-dependent single-particle spectra are calculated by applying a window function with a width of 5.0 fs—corresponding to the pump pulse used in photoemission spectroscopy—so that changes on a half-cycle timescale can be resolved. Since the number of excited carriers is extracted from the photoelectron spectra, it has the same effective time resolution. The sub-cycle emission spectra are evaluated using a window function with a width of 1.5 fs, allowing us to resolve variations within each sub-cycle. Meanwhile, for the half-cycle–resolved emission intensity and phase shown in Figs. 5j, k, and l, we use a window function with a width equal to one half-cycle of the MIR electric field. This provides a temporal resolution on the order of half the period of the driving field.

**Data availability**

The data of this study are available from the corresponding author upon reasonable request.

**Code availability**

The codes are available from the corresponding author upon reasonable request.

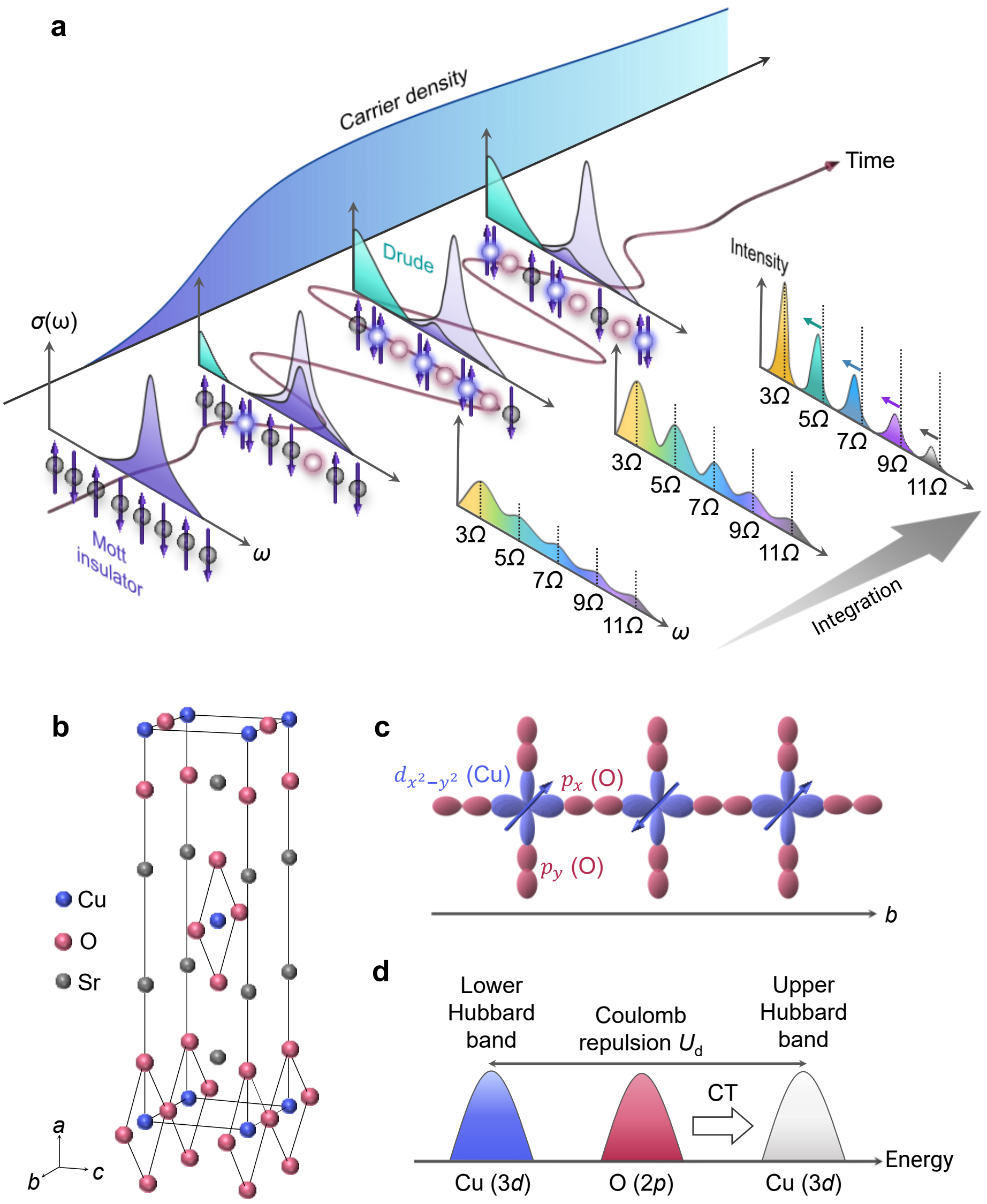

**Fig. 1 | High-harmonic generation (HHG) during melting of a Mott insulating state induced by a mid-infrared (MIR) pulse in a one-dimensional (1D) cuprate, $Sr_2CuO_3$. a**, Schematics of the time characteristics of the electronic-state changes and HHG under a strong MIR electric field. The spectral weight of the optical transition beyond the Mott gap is transferred to the Drude component by doublon and holon creation. Accumulated HH spectra can shift to lower energies with progress in metallisation. **b**, Crystal structure of $Sr_2CuO_3$. **c**, 1D electronic state consisting of the $d_{x^2-y^2}$ orbital of Cu and $p$ orbital of O. **d**, Electronic structure of $Sr_2CuO_3$. The lowest electronic transition is the charge transfer (CT) transition from the $\mathrm{O}-2p$ band to $\mathrm{Cu}-3d$ upper Hubbard band; its energy is denoted by $\Delta_{\mathrm{Mott}}$ in the text.

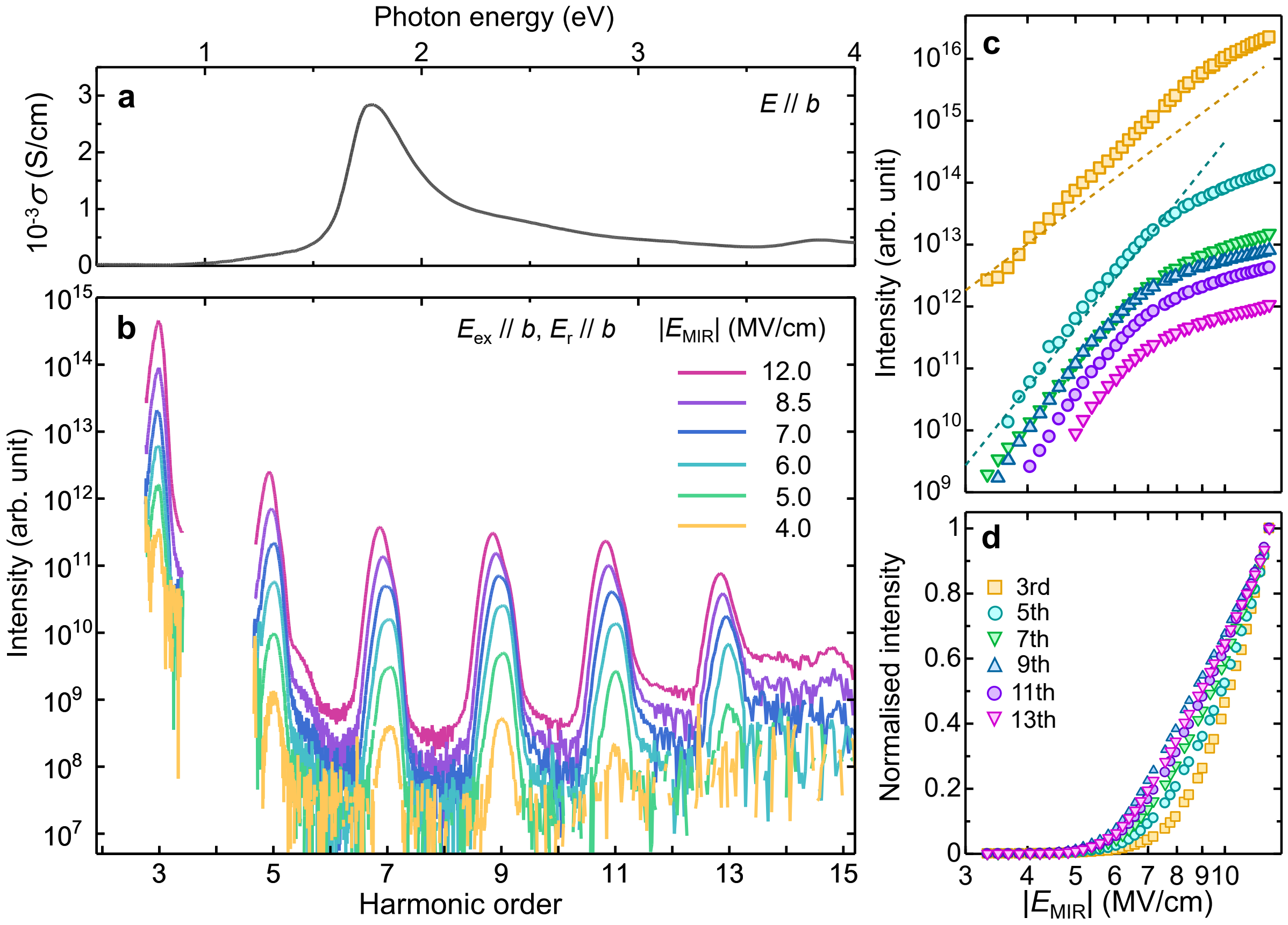

**Fig. 2 | HHG properties as a function of the MIR electric-field amplitude $|E_{\mathrm{MIR}}|$ in $Sr_2CuO_3$. a**, Spectrum of optical conductivity $\sigma(\omega)$ with the electric field of lights, $E$, parallel to the $b$-axis ($E//b$). **b**, Spectra of HHs with the electric field of lights $E_{\mathrm{r}}//b$ obtained by the irradiation of the MIR pump pulse (0.263 eV) with the electric field of lights $E_{\mathrm{ex}}//b$. **c**, **d** $|E_{\mathrm{MIR}}|$ dependence of the integrated intensities of HHs: (**c**) log–log and (**d**) semi-log plots. The yellow (green) broken line in **c** shows the relation $I_{3\mathrm{HG}} \propto |E_{\mathrm{MIR}}|^6$ ($I_{5\mathrm{HG}} \propto |E_{\mathrm{MIR}}|^{10}$).

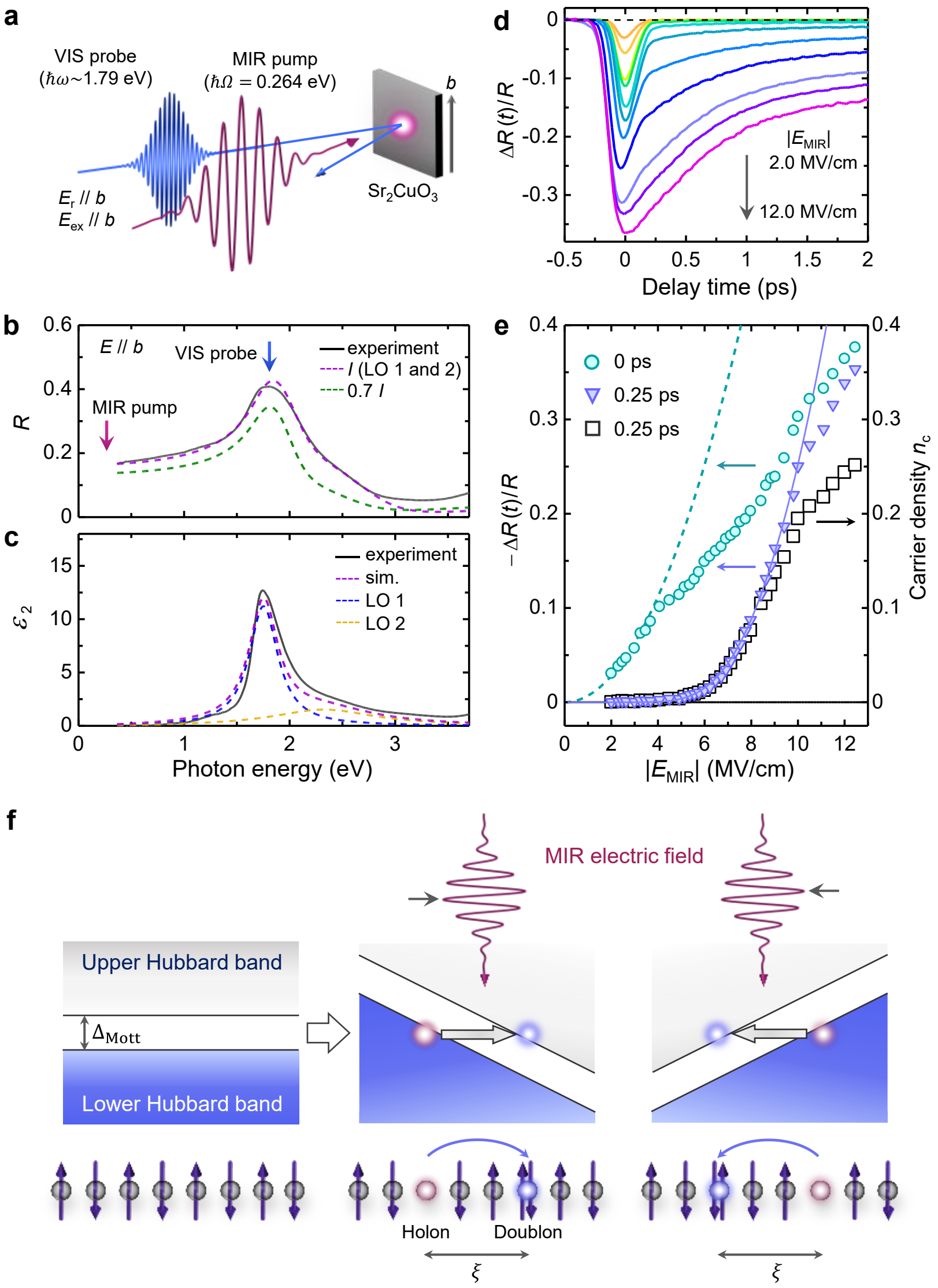

**Fig. 3 | MIR pump-visible reflectivity probe measurements to evaluate the carrier density in $Sr_2CuO_3$. a**, Experimental setup. The arrows show the pump and probe pulse energies. **b, c,** Spectra of the (**b**) reflectivity $R$ and (**c**) imaginary part of the relative dielectric constant, $\varepsilon_2$. The purple broken lines in (**b**) and (**c**) show the spectra simulated using two Lorentz-oscillators (LO 1 and 2) represented by the blue and yellow broken lines, respectively, in **c**. The green broken line in **b** is the simulated spectrum for the intensity reduction ($-\Delta I/I$) required for the total oscillator strength $I$ to be 0.3. **d**, Time characteristics of the reflectivity changes $-\Delta R(t)/R$ for $|E_{\mathrm{MIR}}| = 2 - 12$ MV/cm in 1 MV/cm steps. **e**, $|E_{\mathrm{MIR}}|$ dependence of $-\Delta R(t)/R$ at $t = 0$ and 0.25 ps (circles and triangles, respectively) and the carrier density $n_c$ at $t = 0.25$ ps (squares). The light-blue and purple broken lines show the relations $-\Delta R(t)/R \propto |E_{\mathrm{MIR}}|^2$ and $-\Delta R(t)/R \propto \Gamma$ in equation (1), respectively. **f**, Schematics of doublon–holon pair creations via quantum tunnelling processes by the MIR electric field.

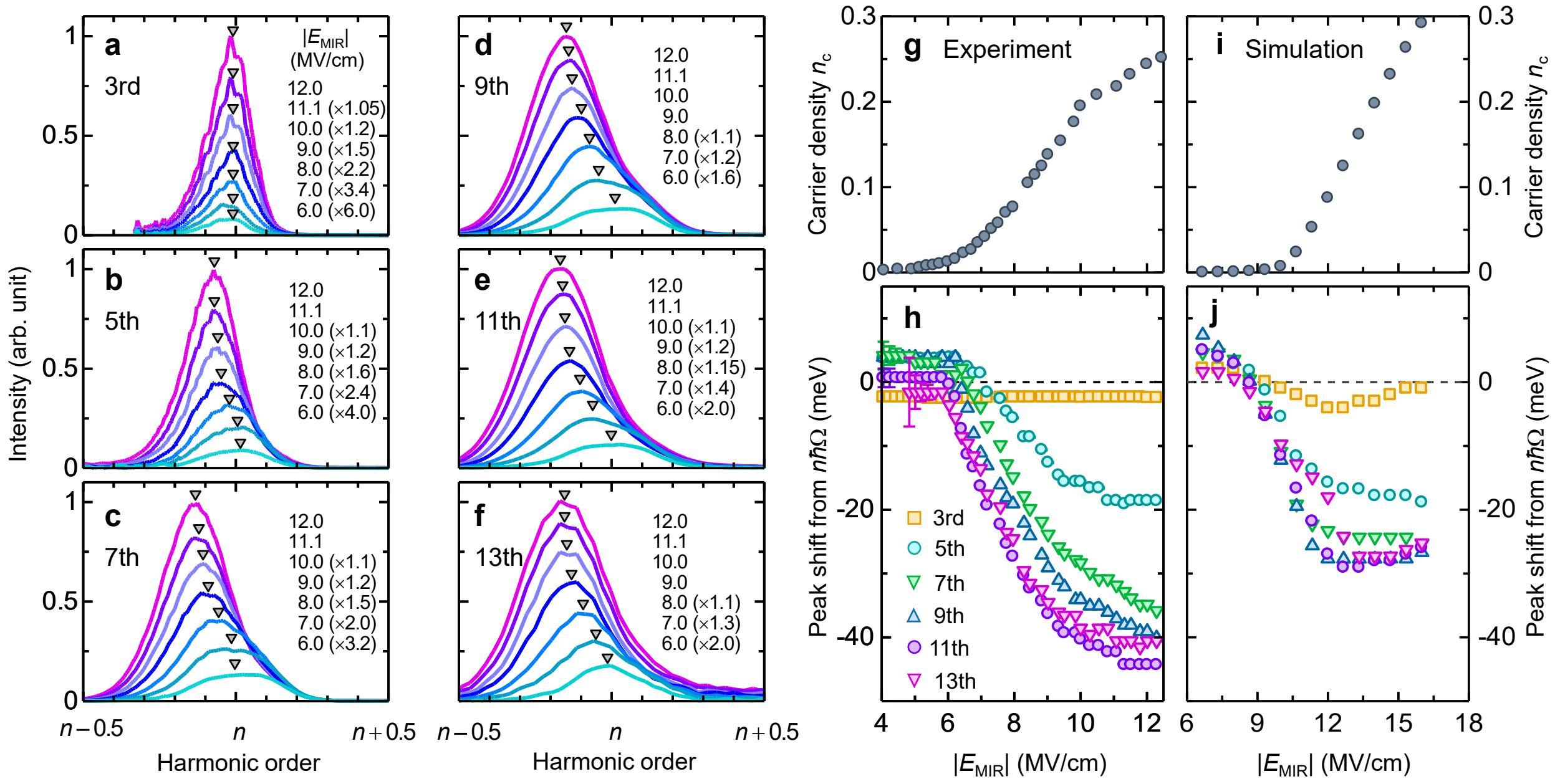

**Fig. 4 | MIR electric-field dependence of the HH spectral shifts in $Sr_2CuO_3$. a–f**, Spectra of 3rd to 13th order harmonics for $|E_{\mathrm{MIR}}| = 6.0 - 12.0$ MV/cm. Triangles show the peak energies (central positions) of each HH spectrum determined by spectral fitting with the Gaussian shape. **g**, $|E_{\mathrm{MIR}}|$ dependence of $n_{\mathrm{c}}$ (the same as the squares in Fig. 3**e**). **h**, $|E_{\mathrm{MIR}}|$ dependence of the HH peak shifts from $n\hbar\Omega$ evaluated from the HH spectra in **a**–**f**. The error bars indicate the standard deviation of the least squares fit. **i**, **j** $|E_{\mathrm{MIR}}|$ dependence of (**i**) $n_{\mathrm{c}}$ after the pulse and (**j**) HH peak shifts from $n\hbar\Omega$ evaluated by the DMFT simulation shown in Fig. 5.

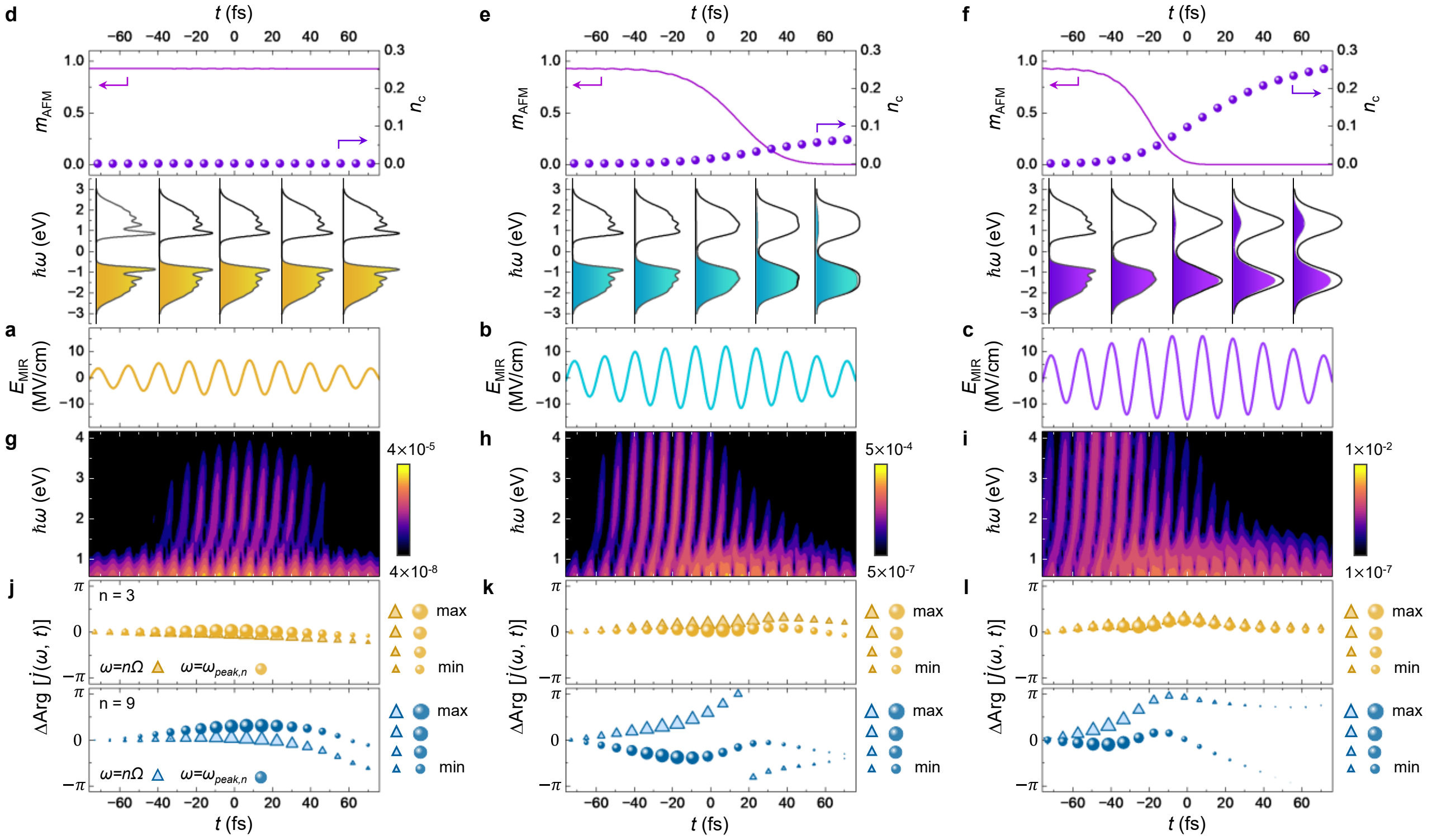

**Fig. 5 | Simulations of HHG in Mott insulators with DMFT. a–c,** Electric-field waveforms of pump pulses with the frequency $\Omega/2\pi = 63$ THz. **d–f,** Upper panels: Antiferromagnetic spin–spin correlation $m_{\mathrm{AFM}}$ and carrier density $n_{\mathrm{c}}$ as a function of $t$. Lower panels: Changes in electronic structure along the electric field of the pump pulse as a function of $t$ for each cycle. Single-particle spectra are shown in solid lines and the filled regimes indicate their occupation. **g–i**, Sub-cycle radiation spectra as a function of $t$. Logarithmic radiation intensity values are indicated by colors defined in the color bar located in the right of each panel. **j–l**, Changes in the radiation phases for each half-cycle, $\Delta\mathrm{Arg}[\dot{j}(\omega, t)]$, at $\omega = n\Omega$ and $\omega = \omega_{\mathrm{peak},n}$ for $n = 3$ and 9. The size of a marker indicates the intensity of the corresponding radiation. $|E_{\mathrm{MAX}}| = 6.7$ MV/cm in **a**, **d**, **g**, and **j**, $|E_{\mathrm{MAX}}| = 12.0$ MV/cm in **b**, **e**, **h**, and **k**, and $|E_{\mathrm{MAX}}| = 16.0$ MV/cm in **c**, **f**, **i**, and **l**.

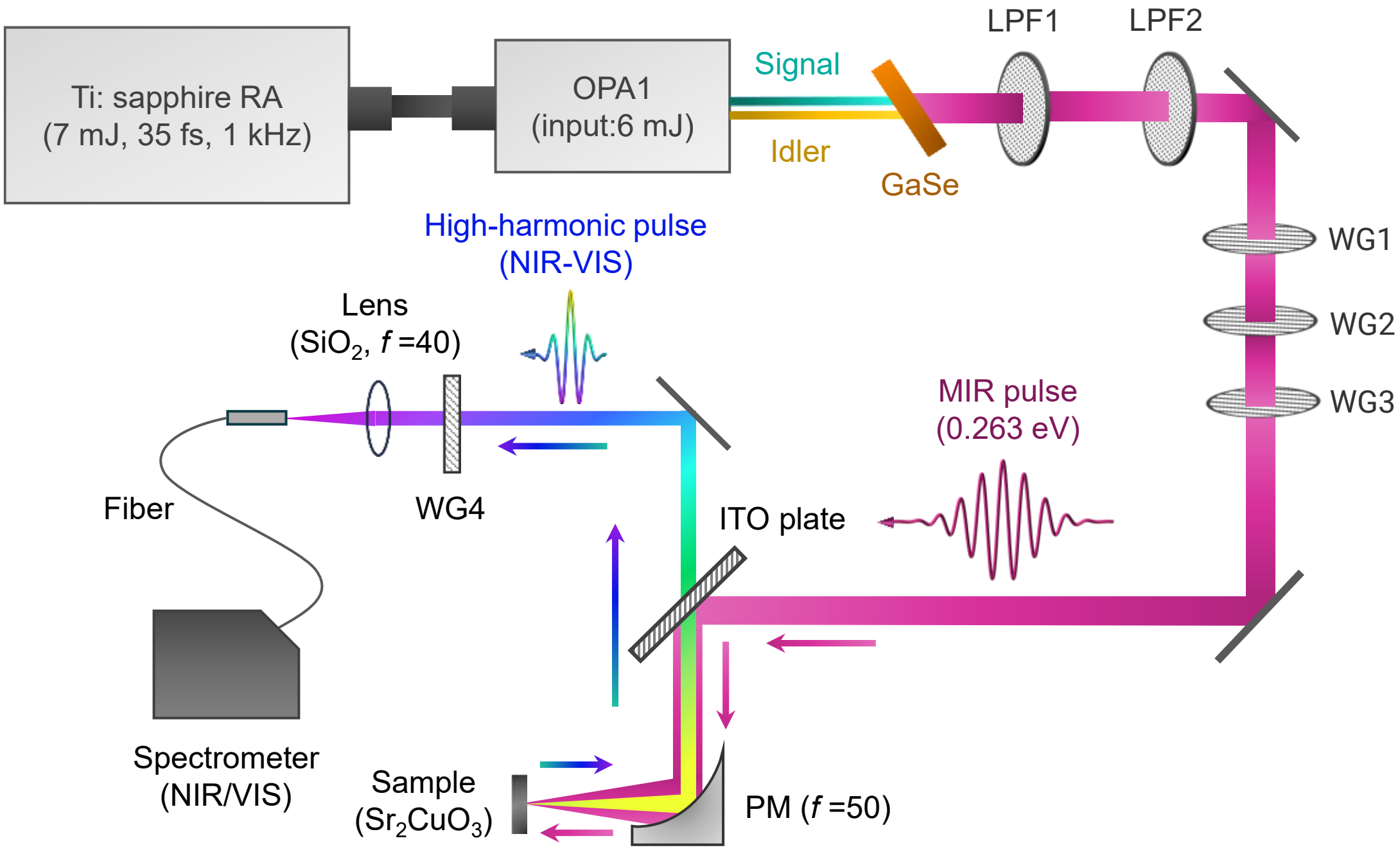

**Extended Data Fig. 1: Schematic diagram of HH spectrum measurements.** LPF1 and LPF2 are long-pass filters which are transparent only at the wavelength longer than 3000 and 3500 nm, respectively. PM is a parabolic mirror, the focus length of which is expressed by $f$ (mm). WG1-3 and WG4 are wire-grid polarizers for the MIR region and NIR-VIS region, respectively.

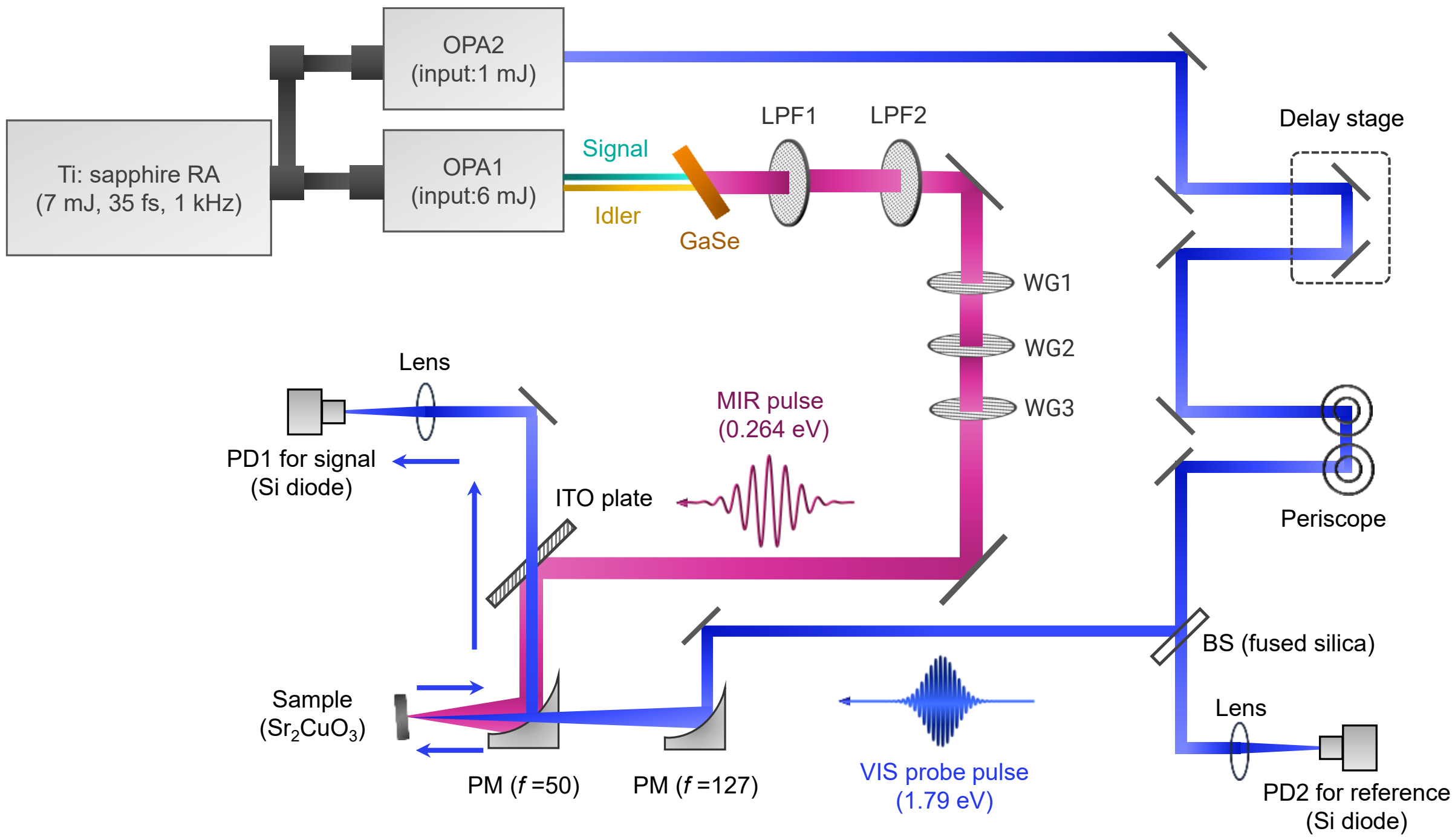

**Extended Data Fig. 2: Schematic diagram of MIR pump-VIS reflectivity probe measurements.** LPF1 and LPF2 are long-pass filters which are transparent only at the wavelength longer than 3000 and 3500 nm, respectively. PM is a parabolic mirror, the focus length of which is expressed by $f$ (mm). WG1-3 are wire-grid polarizers for the MIR region. BS is a beam splitter. PD1 and PD2 are photodetectors (Si diode).

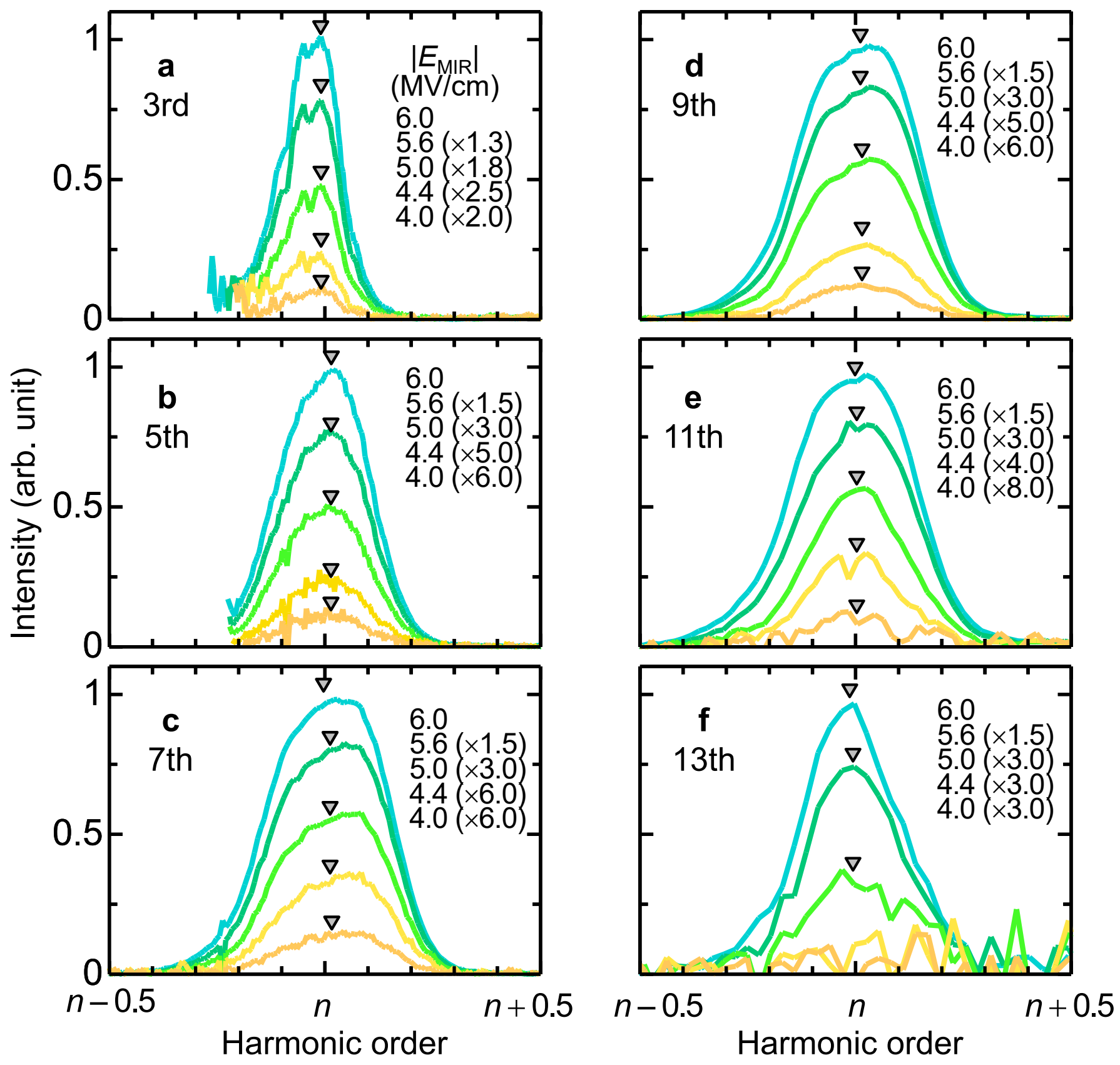

**Extended Data Fig. 3: MIR electric-field dependence of spectral shifts of HHs in $Sr_2CuO_3$ at the lower electric fields.** Spectra of 3rd-13th harmonics for $|E_{\mathrm{MIR}}|$ of 4.0 to 6.0 MV/cm. Triangles show peak energies (central positions) of each HH spectrum determined by the spectral fittings with the Gaussian shape.

## Supplementary information:

# High-harmonic generation during ultrafast melting of a Mott insulating state

Ryohei Ikeda[1,2], Yuta Murakami[3,4], Daiki Sakai[1], Tatsuya Miyamoto[1,5], Toshimitsu Ito[6], Hiroshi Okamoto[1,7]

[1]*Department of Advanced Materials Science, University of Tokyo, Chiba 277-8561, Japan*
[2] *Institute of Solid states Physics, University of Tokyo, Chiba 277-8561, Japan*
[3]*Institute for Materials Research, Tohoku University, Sendai 980-8577, Japan*
[4]*Center for Emergent Matter Science, RIKEN, Wako, Saitama 351-0198, Japan*
[5]*Department of Engineering, Nagoya Institute of Technology, Nagoya, 466-8555, Japan*
[6]*National Institute of Advanced Industrial Science and Technology, Tsukuba 305-8565, Japan*
[7]*Department of Physics, Nihon University, Tokyo 101-8308, Japan*

**Table of contents**

## S1 Evaluation of temporal width and maximum electric field of MIR pump pulses

Figures S1(a) and (b) show the spectra of the MIR pulses used for pump-probe reflectivity (PPR) and HHG measurements, respectively. The central photon energies of the MIR pulses were 0.264 and 0.263 eV for the PPR and HHG measurements, respectively, which are almost the same. The full widths at half maximum (FWHM) of the MIR pulses were 24 meV and 22 meV for the PPR and HHG measurements, respectively.

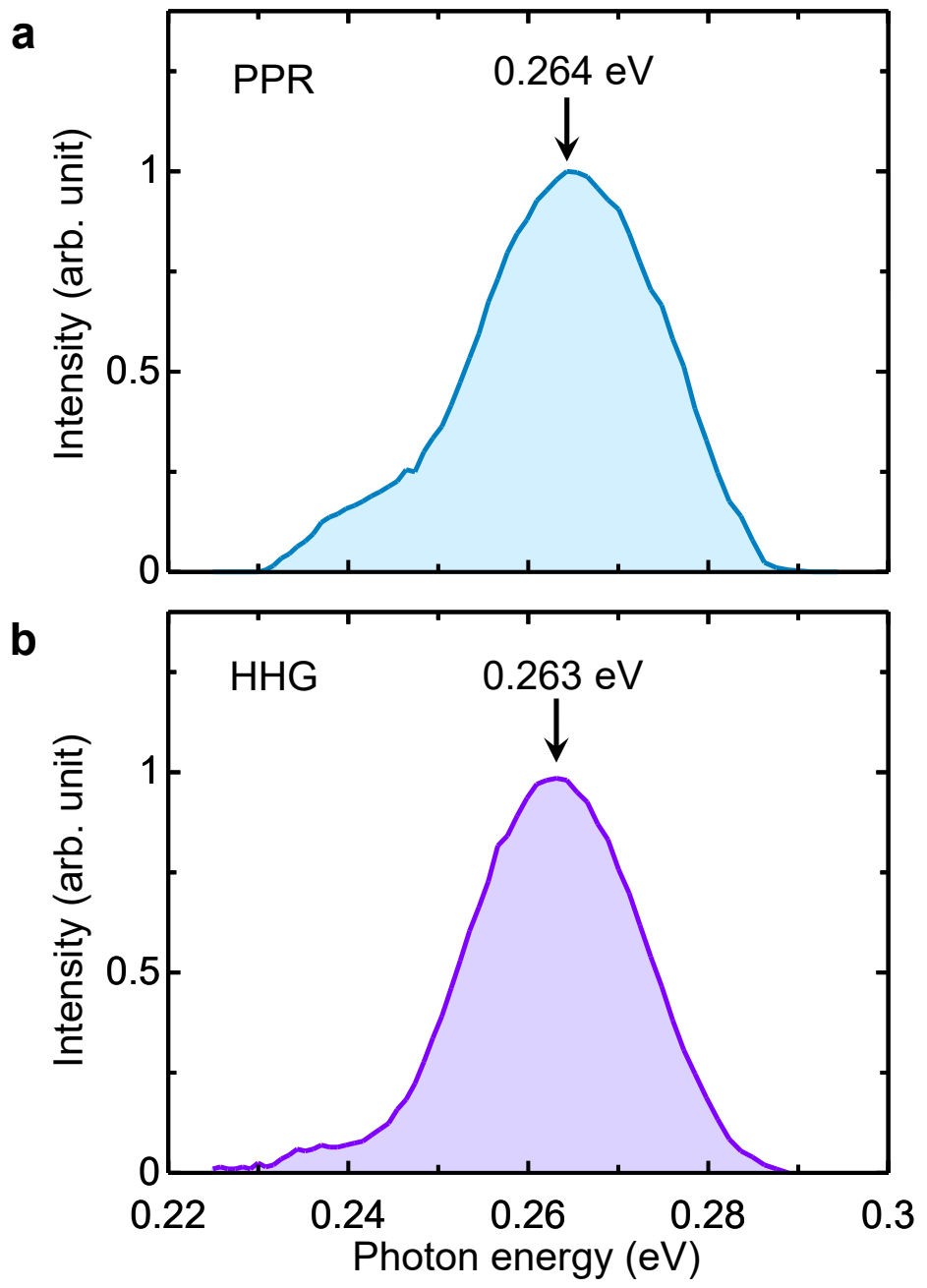

**Fig. S1 | Spectra of MIR pulses.** (**a**) PPR and (**b**) HHG measurements.

Although estimating the temporal width of an MIR pulse is generally difficult, it is possible by using the time characteristic of the reflectivity change $-\Delta R/R(t)$ under weak excitation (Fig. 3d), which can be regarded as a coherent response, obtained from PPR measurements. As a typical example of such a coherent response, Fig. S2(a) shows the time characteristic of $-\Delta R/R(t)$ for $E_{\mathrm{MIR}} = 3.0\ \mathrm{MV/cm}$. Next, we explain this response. In 1D Mott insulators, a one-photon-allowed excitonic state with odd-parity, $|\varphi_{\mathrm{o}}\rangle$, is the lowest excited state and a one-photon-forbidden excitonic state with even-parity, $|\varphi_{\mathrm{e}}\rangle$, is located just above $|\varphi_{\mathrm{o}}\rangle$[1-4], as shown on the left side of Fig. S2(b). When an oscillating electric field of frequency $\Omega$, which is larger than the frequency difference

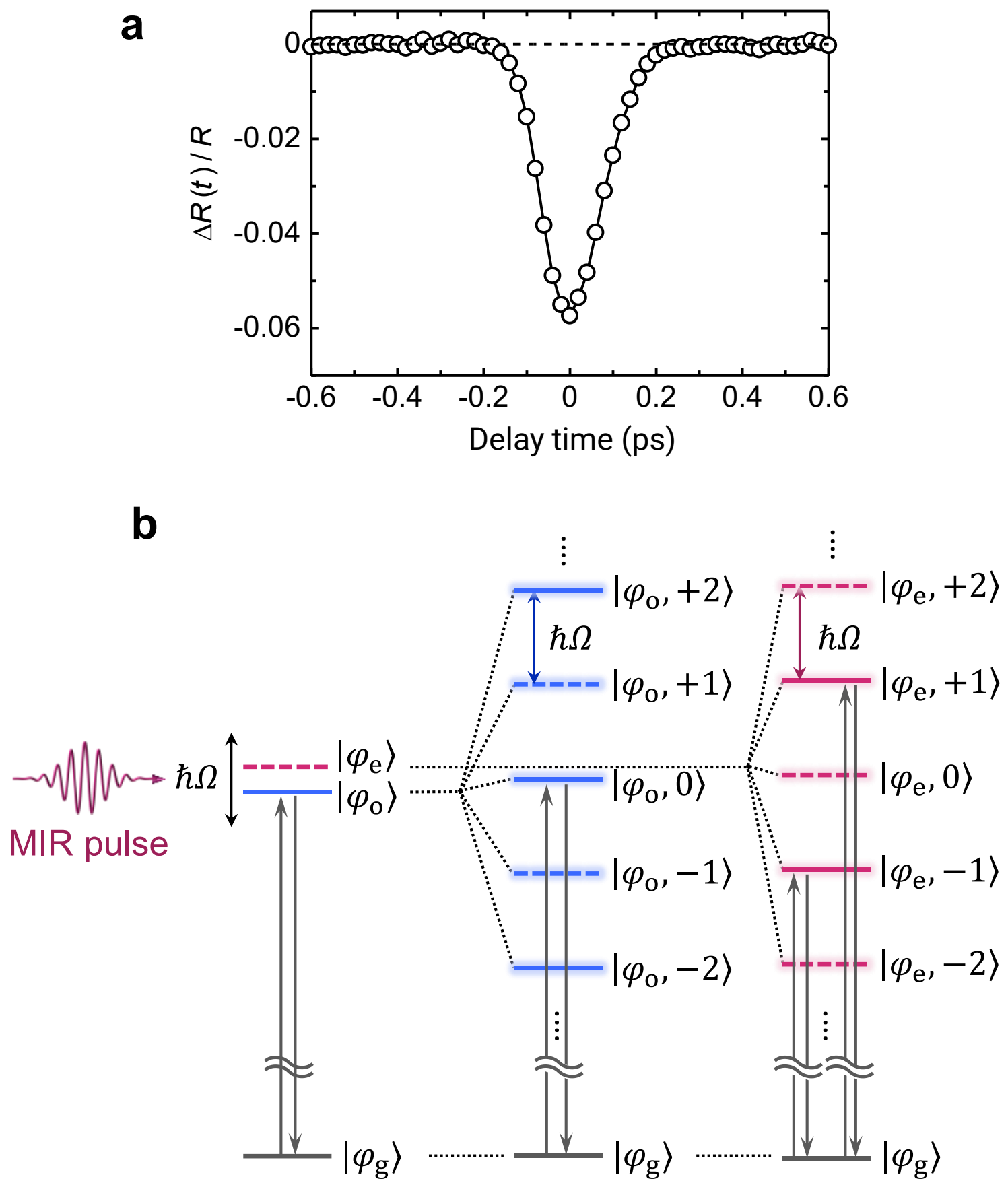

**Fig. S2 | Coherent responses associated with excitonic Floquet states under MIR electric fields. a,** The time characteristic of $-\Delta R/R(t)$ for the amplitude of the MIR pump pulse, $|E_{\mathrm{MIR}}| = 3.0\mathrm{MV/cm}$, extracted from Fig. 3d in $Sr_2CuO_3$ in the main paper. The photon energy $\hbar\Omega$ of the MIR pump light is 0.263 eV. The polarizations of the MIR pump and VIS probe lights are both parallel to the $b$-axis. **b,** Energy-level structures of the excitonic Floquet states and related optical processes in 1D Mott insulators. The left panel shows the original excitonic states $|\varphi_{\mathrm{o}}\rangle$ and $|\varphi_{\mathrm{e}}\rangle$. The middle and right panels show the excitonic Floquet states of the odd-parity exciton $|\varphi_{\mathrm{o}}, m\rangle$ and even-parity exciton $|\varphi_{\mathrm{e}}, m\rangle$, respectively, under an MIR electric field of frequency $\Omega$. Vertical arrows indicate linear optical processes involving the incident and emitted light.

between $|\varphi_{\mathrm{o}}\rangle$ and $|\varphi_{\mathrm{e}}\rangle$, is applied to this state, $|\varphi_{\mathrm{o}}\rangle$ and $|\varphi_{\mathrm{e}}\rangle$ are converted to $|\varphi_{\mathrm{o}}, 0\rangle$ and $|\varphi_{\mathrm{e}}, 0\rangle$, respectively, and move slightly closer together[5]. Furthermore, the sidebands $|\varphi_{\mathrm{o}}, \pm m\rangle$ and $|\varphi_{\mathrm{e}}, \pm m\rangle$ are created at a frequency $(\pm m\Omega)$ apart from $|\varphi_{\mathrm{o}}, 0\rangle$ and $|\varphi_{\mathrm{e}}, 0\rangle$, respectively (the right part of Fig. S2(b)), where $m$ is a natural number. This state under an oscillating electric field is generally called the Floquet state[6-8]. In this

situation, $|\varphi_o, 0\rangle$, $|\varphi_o, \pm 2m\rangle$, and $|\varphi_e, \pm(2m-1)\rangle$ are one-photon allowed states. Considering only third-order optical nonlinearity, the oscillator strength of the transition from the ground state $|\varphi_g\rangle$ to $|\varphi_o\rangle$ without an MIR electric field is transferred not only to the $|\varphi_g\rangle \rightarrow |\varphi_o, 0\rangle$ transition but also to the $|\varphi_g\rangle \rightarrow |\varphi_e, \pm 1\rangle$ transitions under an MIR electric field[5]. As a result, the reflectivity around the peak corresponding to the $|\varphi_g\rangle$ to $|\varphi_o\rangle$ transition decreases. Such excitonic Floquet state formation is a type of coherent response that occurs only when the MIR pulse is irradiating[5,9-11], resulting in a pulsed signal reflecting its coherent nature, as shown in Fig. S2(a).

Therefore, the time characteristic of $-\Delta R/R(t)$ shown in Fig. S2(a) should be proportional to the convolution integral of the intensity profiles of the MIR pump pulse and visible probe pulse. In addition, as described in Section S3, the observed $-\Delta R/R(t)$ signal exhibits a Gaussian shape centred at the time origin. In this case, the temporal width $t_S$ (FWHM) of this coherent response, $-\Delta R/R(t)$, can be expressed using the temporal widths $t_{MIR}$ and $t_{VIS}$ (FWHM) of the MIR pump pulse and visible probe pulse, respectively, as follows:

$$t_S = ({t_{MIR}}^2 + {t_{VIS}}^2)^{\frac{1}{2}}. \quad \text{(S1)}$$

The $t_{VIS}$ of the probe pulse was ~55 fs, whereas the $t_S$ of the reflectivity-change signal (Fig. 3d and Fig. S2(a)) at $|E_{MIR}| = 3.0$ MV/cm was ~180 fs. From these values, the $t_{MIR}$ of the MIR pulse used for PPR measurements was estimated to be ~170 fs. Using the same method, the $t_{MIR}$ of the MIR pulse used for HHG measurements was estimated to be ~220 fs.

The maximum electric-field amplitude $|E_{MIR}|$ of an MIR pulse in each measurement was calculated from the photon energy $\hbar\Omega$, temporal width $t_{MIR}$, fluence $F_s$ of the MIR pulse, and spot size $l_S$ (FWHM) at the sample position of the MIR pulse. In the PPR and HHG measurements, $l_S = 45$ and $32.5$ μm, respectively, and the maximum powers of the MIR pulses are 1.51 and 0.79 μJ, respectively, from which the $|E_{MIR}|$ values are estimated to be 13.7 and 12.3 MV/cm, respectively. An estimation error for each electric field amplitude $|E_{MIR}|$ is within ± 0.1 MV/cm at any electric field strength. The spot size of the probe pulse in PPR measurements (23.6 μm) was sufficiently small compared with that of the pump pulse (45 μm).

## S2 Polarisation dependence of HH spectra

Figures S3(a–c) show the HH spectra with the light polarisation $E_r$ parallel and perpendicular to the $b$-axis ($E_r//b$ and $E_r \perp b$, respectively) in $Sr_2CuO_3$ on excitation with an MIR pulse (0.263 eV and $|E_{MIR}| = 12.3$ MV/cm) with the light polarisation $E_{ex}$ parallel or perpendicular to the $b$-axis ($E_{ex}//b$ or $E_{ex} \perp b$, respectively). When the HH intensity for $E_{ex}//b$ and $E_r//b$ is expressed as $I_{pp}$, that for $E_{ex}//b$ and $E_r \perp b$, $I_{ps}$, is roughly equal to $10^{-2}I_{pp}$, and that for $E_{ex} \perp b$ and $E_r//b$, $I_{sp}$, is smaller than $10^{-2}I_{pp}$, except for the 3rd order harmonic. The HH intensity for $E_{ex} \perp b$ and $E_r \perp b$, $I_{ss}$, is smaller than $I_{ps}$ and $I_{sp}$ (not shown). These results are consistent with the strong one-dimensionality of the electronic state of $Sr_2CuO_3$.

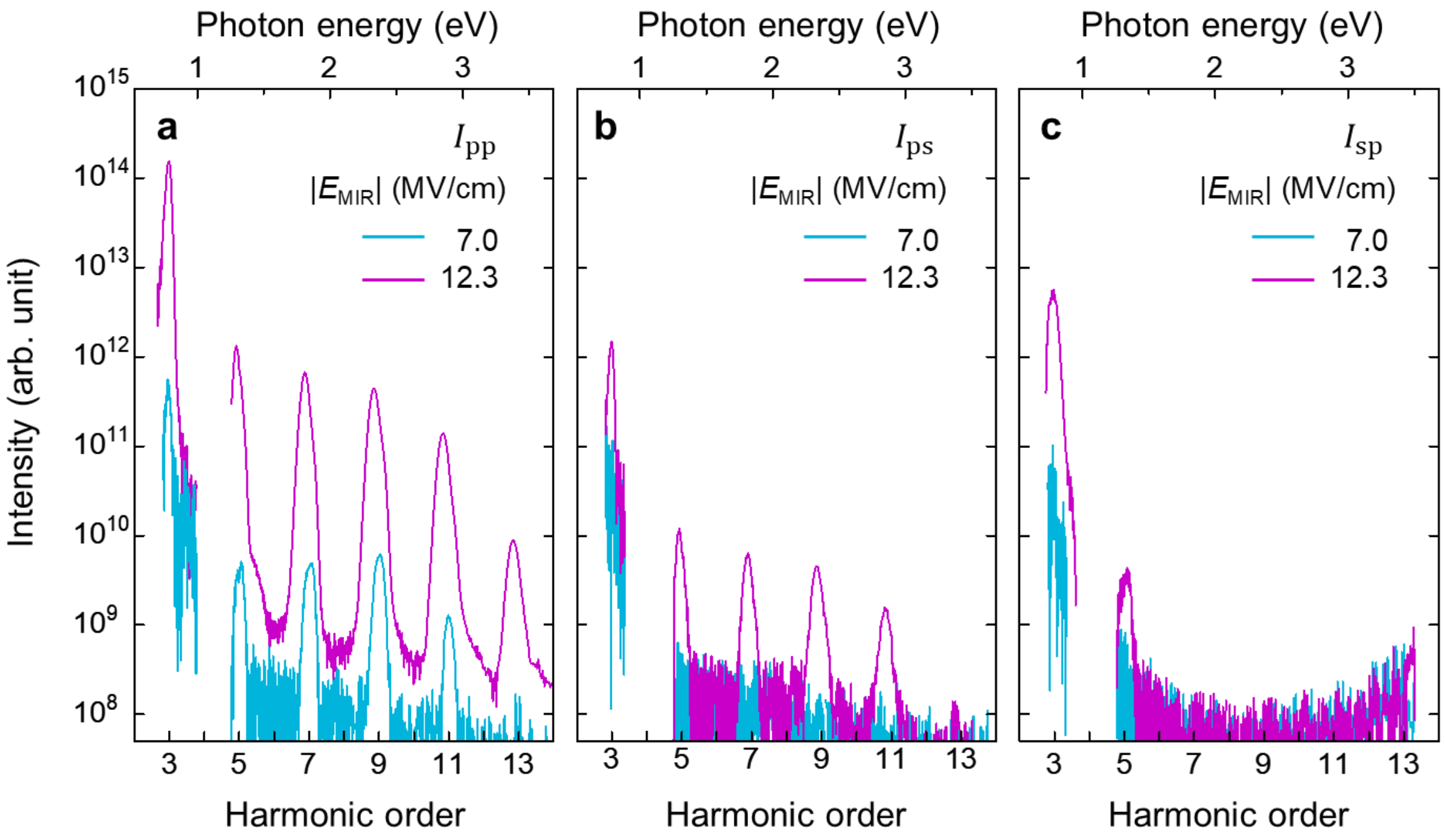

**Fig. S3 | Polarization dependence of HH spectra in $Sr_2CuO_3$.** The photon energy $\hbar\Omega$ is 0.263 eV, and the electric field amplitude $|E_{MIR}|$ is 12.3 or 7.0 MV/cm. (**a**) $I_{pp}$: $E_{ex}//b$ and $E_r//b$, (**b**) $I_{ps}$: $E_{ex}//b$ and $E_r \perp b$, and (**c**) $I_{sp}$: $E_{ex} \perp b$ and $E_r//b$. $E_{ex}$ and $E_r$ are the polarization of the MIR pump and HH light, respectively.

## S3 Fitting analysis of transient reflectivity changes induced by MIR pump pulses

In this section, we discuss the analysis of the time characteristics of the reflectivity changes $\Delta R(t)/R$ induced by an MIR pulse (shown in Fig. 3d of the main text). Figures S4(a), (b), and (c, d) show the time characteristics of $\Delta R(t)/R$ for the MIR pulse with $|E_{\mathrm{MIR}}|$ =3.0, 7.0, and 12.0 MV/cm. As described in Section S1, in the $\Delta R(t)/R$ signal induced by the MIR pulse with $|E_{\mathrm{MIR}}|$ =3.0 MV/cm, only the coherent response due to the 3rd order optical nonlinearity appears. This signal can be fitted with the following Gaussian-type function, as indicated by the light blue line in Fig. S4(a):

$$\frac{\Delta R(t)}{R} = -A_0 \exp\left[-\left(\frac{t}{\tau_0}\right)^2\right]. \tag{S2}$$

$\tau_0$ is 108 fs, which corresponds to $t_{\mathrm{S}}\sim$180 fs, as mentioned in Section S1.

As described in the main text, at $|E_{\mathrm{MIR}}|$ =7.0 and 12.0 MV/cm, a component with a finite decay time due to generated carriers appears in the $\Delta R(t)/R$ signals. When carriers are annihilated by recombination, the signals due to the increase in temperature are superimposed. According to a previous study, the reflectivity at 1.79 eV, where $\Delta R(t)/R$ is measured, decreases with increasing temperature[2]. Following the previously reported analysis of photoinduced transient reflectivity changes in Mott insulators, we assume that the time characteristic of the carrier recombination is expressed as an exponential function with a decay time of $\tau_1$; the time characteristics of the increase in temperature are expressed by a term including two exponential functions representing its rise and decay[12]. The rise and decay times are $\tau_1$ and $\tau_2$, respectively. In this case, $\Delta R(t)/R$ can be expressed as follows:

$$\frac{\Delta R(t)}{R} = -A_0 \exp\left[-\left(\frac{t}{\tau_0}\right)^2\right] - \int_{-\infty}^{t} A_1 \exp\left(-\frac{t-t'}{\tau_1}\right) \frac{1}{\sqrt{\pi}\tau_0} \exp\left[-\left(\frac{t'}{\tau_0}\right)^2\right] dt'$$

$$- \int_{-\infty}^{t} A_2 \left[1 - \exp\left(-\frac{t-t'}{\tau_1}\right)\right] \left[\exp\left(-\frac{t-t'}{\tau_2}\right)\right] \frac{1}{\sqrt{\pi}\tau_0} \exp\left[-\left(\frac{t'}{\tau_0}\right)^2\right] dt'. \tag{S3}$$

The first term represents the coherent response ($\tau_0 = 0.108$ ps) expressed by equation (S2), the second term represents the reduction in reflectivity due to electric-field induced carriers that disappear with the decay time of $\tau_1$, and the third term represents the reduction in reflectivity due to the increase in temperature with the rise time of $\tau_1$ and decay time of $\tau_2$. The time characteristics of $\Delta R(t)/R$ at $|E_{\mathrm{MIR}}| = 7.0$ and 12.0 MV/cm are well reproduced by this equation, as shown by the blue lines in Figs. S4(b–d). The three components represented by the first, second, and third terms are indicated by light blue, pink, and yellow lines, respectively. The parameters obtained from this fitting analysis are listed in Table S1 along with the results for $|E_{\mathrm{MIR}}| =$ 3.0 MV/cm. For $|E_{\mathrm{MIR}}| = 7.0$ and 12.0 MV/cm, $\tau_1$ was 0.60 and 0.77 ps, respectively, and $\tau_2$ was 150 ps. The magnitude of the coherent response, $A_0$, decreased considerably at $|E_{\mathrm{MIR}}| =$12.0 MV/cm compared to that at $|E_{\mathrm{MIR}}| = 7.0$ MV/cm, likely owing to the disruption of coherence caused by carrier generation.

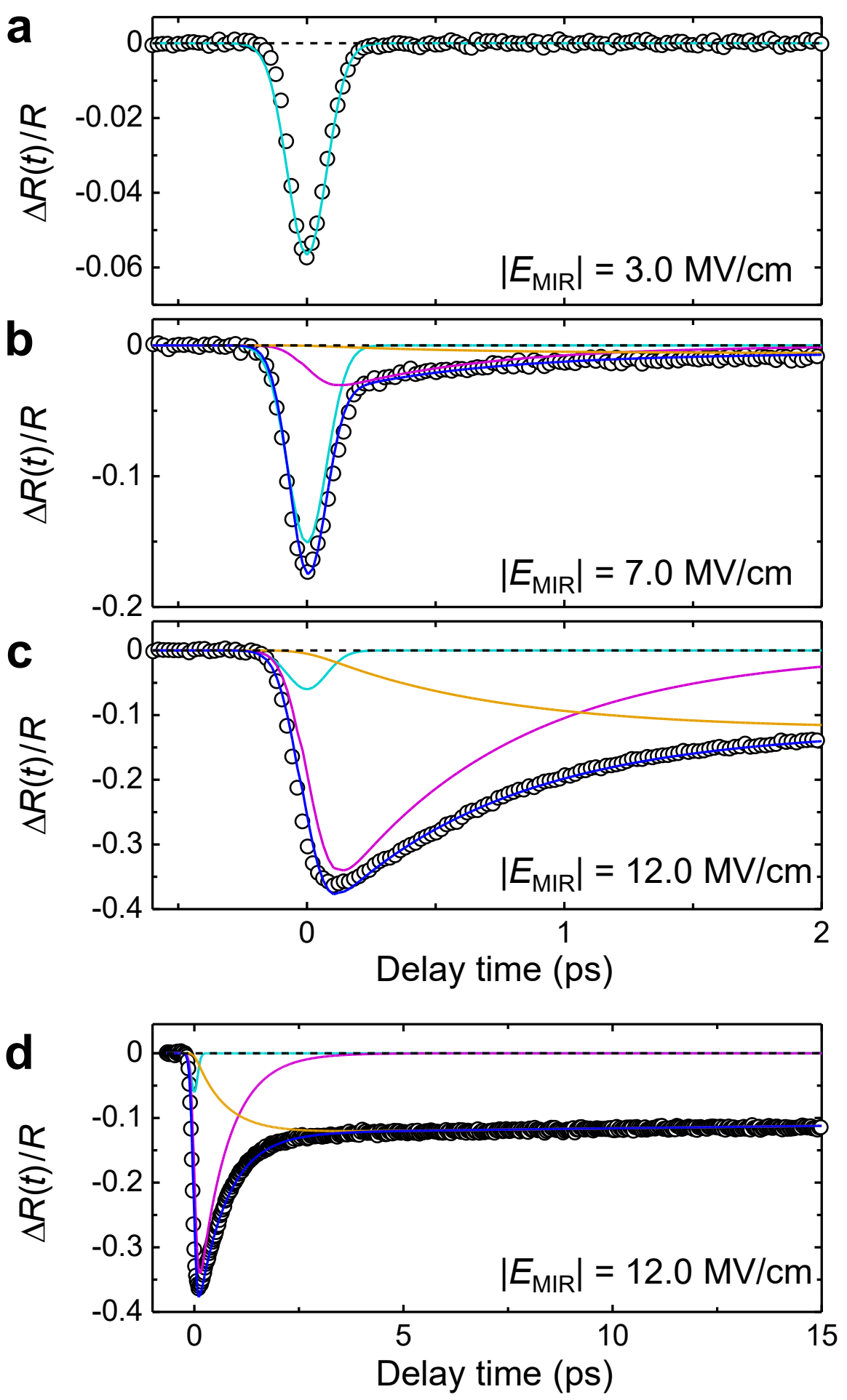

**Fig. S4 | Analyses of the time characteristics of $-\Delta R(t)/R$ in $Sr_2CuO_3$.** (**a**) $|E_{\mathrm{MIR}}| = 3.0\mathrm{MV/cm}$, (**b**) $|E_{\mathrm{MIR}}| = 7.0\mathrm{MV/cm}$, and (**c, d**) $|E_{\mathrm{MIR}}| = 12.0\mathrm{MV/cm}$. The photon energy of the MIR pump light, $\hbar\Omega$, is 0.264 eV. $E_{\mathrm{ex}}$ and $E_{\mathrm{r}}$ are both parallel to the $b$-axis. Data marked by open circles are extracted from Fig. 3d in the main text. Light-blue, pink, and yellow lines indicate the components of the first, second, and third terms in equation (S3). The blue lines in (**b–d**) indicate the sum of three components corresponding to the three terms in equation (S3).

**Table S1 | Fitting parameters of $-\Delta R(t)/R$ in $Sr_2CuO_3$.**

| $\lvert E_{MIR}\rvert$ (MV/cm) | 3.0 | 7.0 | 12.0 |
|---|---|---|---|
| $A_0$ | 0.033±0.002 | 0.092±0.004 | 0.035±0.05 |
| $\tau_0$ (ps) | 0.108±0.005 | 0.108±0.005 | 0.108±0.005 |
| $A_1$ | 0 | 0.04±0.004 | 0.43±0.014 |
| $\tau_1$ (ps) | – | 0.6±0.05 | 0.7±0.03 |
| $A_2$ | 0 | 0.006±0.0005 | 0.124±0.003 |
| $\tau_2$ (ps) | – | 150±30 | 150±30 |

**S4 Evaluation of the carrier density generated by MIR pump pulses**

In this section, we discuss the procedure followed to estimate the carrier density $n_c$ from the reflectivity change $\Delta R(t)/R$ mentioned in the previous section in $Sr_2CuO_3$. The $-\Delta R(t)/R$ at $t = 0.25$ ps, $-\Delta R(0.25\text{ps})/R$, when the coherent response completely disappears, is mainly attributable to generated carriers. Therefore, the magnitude of $-\Delta R(0.25\text{ps})/R$ can be used to estimate $n_c$. More precisely, since $-\Delta R(0.25\text{ps})/R$ includes the signal due to the temperature increase represented by the yellow line in Figs. S4(b–d), the net amount of carriers can be attributed to 0.91 of the $-\Delta R(0.25\text{ps})/R$ signal for $\lvert E_{MIR}\rvert$ =7.0 MV/cm and 0.9 of the $-\Delta R(0.25\text{ps})/R$ signal for $\lvert E_{MIR}\rvert =$ 12 MV/cm. Notably, it is reasonable to estimate the value of $n_c$ from the maximum absolute value of $-\Delta R(t)/R$ represented by the pink line, which is 1.13 of $-\Delta R(0.25\text{ps})/R$ for $\lvert E_{MIR}\rvert$ =7.0 MV/cm and 1.13 of $-\Delta R(0.25\text{ps})/R$ for $\lvert E_{MIR}\rvert =$ 12 MV/cm. Since these corrections cancel out, it is reasonable to estimate $n_c$ from the magnitude of $-\Delta R(0.25\text{ps})/R$ itself.

Next, it is necessary to convert the magnitude of $-\Delta R(0.25\text{ps})/R$ to the change in the imaginary part of the dielectric constant, $\varepsilon_2$, showing the absorption. Regarding the $\varepsilon_2$ spectrum of $Sr_2CuO_3$, the following is known from the previous results of linear and third-order nonlinear optical spectroscopy and their theoretical analyses[1,2,13,14].

· The spectral intensity in the 1-3 eV region of $\varepsilon_2(\omega)$ can be considered to originate from the transitions from the O 2*p* band to the Cu 3*d* upper Hubbard band. The peak around ~1.8 eV is due to the exciton absorption, while the spectrum at the higher energies is due to the absorption by free doublon (D)-holon (H) pairs.

· The exciton absorption is broad, making the exciton peak and edge of the absorption by free D-H pairs appear continuous. Furthermore, the latter absorption tends to become stronger near its absorption edge. This is a common feature also observed in the near-infrared to visible $\varepsilon_2(\omega)$ spectra of the other 1D CT insulators of halogen-bridged Ni compounds, the representative 1D Mott insulator of the organic molecular compound, bis(ethylenedithio)tetrathiafulvalene-difluorotetracyano-quinodimethane (ET-$F_2$TCNQ), and the representative 2D CT insulators of $Nd_2CuO_4$, $La_2CuO_4$, and $Sr_2CuO_2Cl_2$ (refs. 3,4,15).

Based upon these facts, these $\varepsilon_2$ spectra are often analyzed as consisting of the exciton absorption and free D-H pair absorption, represented by two Lorentzian oscillators, in the Hubbard model. Indeed, these analyses well describe the fundamental features of both linear and third-order nonlinear spectra of the complex dielectric function $\tilde{\varepsilon}(\omega)$ and optical conductivity $\tilde{\sigma}(\omega)$[3,4,15]. In the present study, therefore, we assumed that $\tilde{\varepsilon}(\omega)$ consisted of two Lorentz terms, LO1 and LO2, in $Sr_2CuO_3$:

$$\tilde{\varepsilon}(\omega) = \varepsilon_\infty + \frac{S_1}{{\omega_1}^2 - \omega^2 - i\gamma_1\omega} + \frac{S_2}{{\omega_2}^2 - \omega^2 - i\gamma_2\omega} \tag{S4}$$

$S_1$ ($S_2$) is a parameter representing the transition intensity, $\omega_1$ ($\omega_2$) is the central frequency, $\gamma_1$ ($\gamma_2$) is the damping constant in the transition to excitons (continuum), and $\varepsilon_\infty$ is the dielectric constant at high frequencies. The $R(\omega)$ and $\varepsilon_2(\omega)$ spectra shown by the black solid lines in Fig. S5(a) and (b), respectively, are almost reproduced by $\tilde{\varepsilon}(\omega)$ in equation (S4), as indicated by broken purple lines in both figures. The $\varepsilon_2$ spectra of the two Lorentz components LO1 and LO2 are shown by blue and yellow solid lines, respectively, in Fig. S5(c). These results are presented in Fig. 3b, c in the main text, and these parameters are listed in Table S2.

To evaluate the carrier density $n_c$, we should estimate how much the peak reflectivity changes when carriers are generated by an MIR pulse. This is possible in the following way. In a carrier doping to CT insulators and Mott insulators, it is crucial that the sum rule for oscillator strength holds. Indeed, the previous study showed that when hole carriers are introduced to a typical CT insulator of 2D cuprate, $La_2CuO_4$, via the element substitution, the effective electron number obtained by integrating the spectral intensity from zero energy up to ~3 eV (corresponding approximately to the upper edge of the CT absorption) remains nearly constant[16]. Furthermore, when another typical CT insulator of the bromine-bridged Ni compound, $[Ni(chxn)_2Br]Br_2$ (chxn: cyclohexandiamine), is excited with a femtosecond laser pulse beyond the gap, it was shown that the sum rule also holds in the energy region including both the Drude component and CT-gap transition[17].

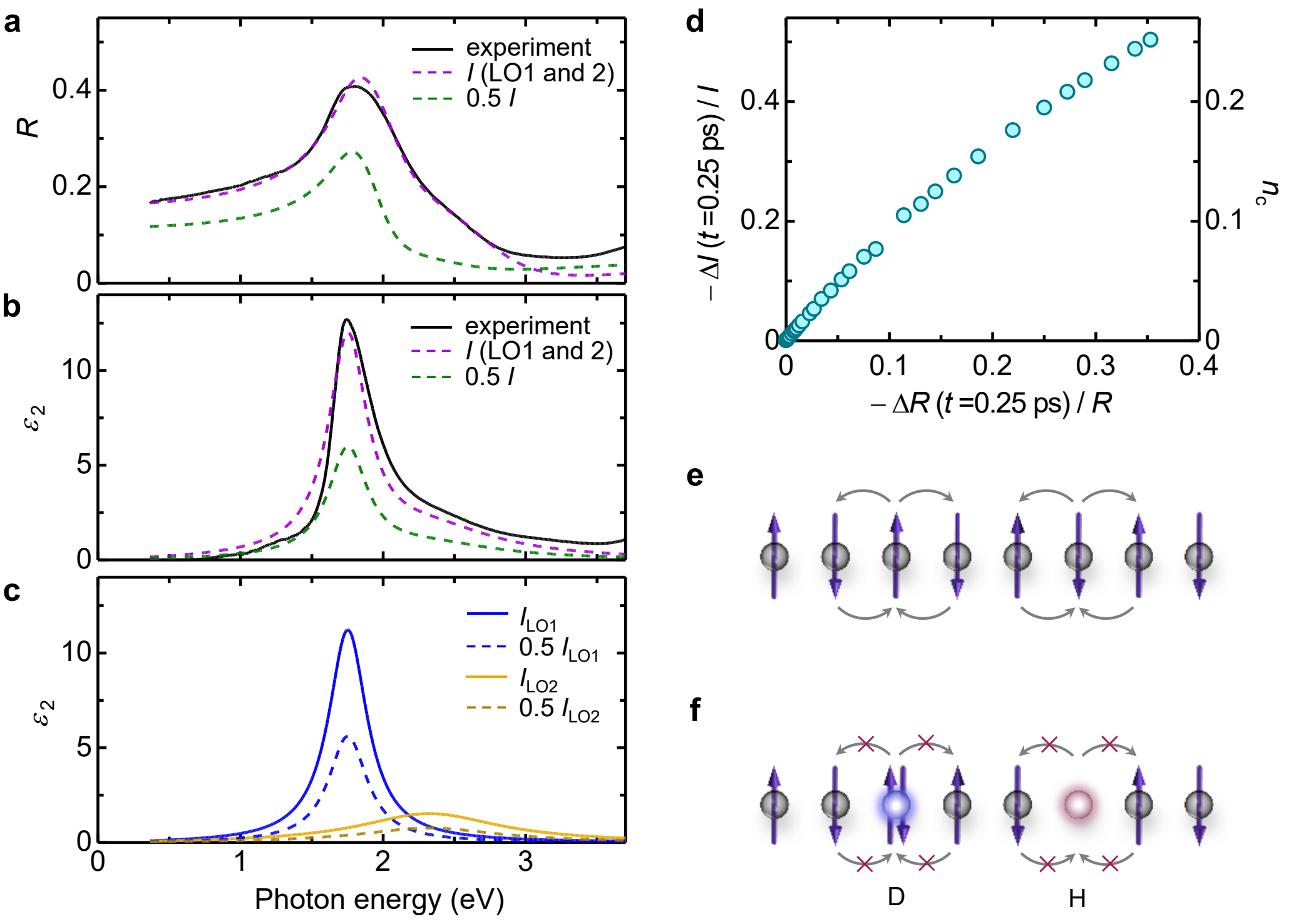

**Fig. S5 | Estimation of the carrier density $n_c$ generated by the MIR pulse in $Sr_2CuO_3$. a,** Reflectivity ($R$) spectra. The black solid and purple broken lines show the experimental result and curve fitted with two Lorentz oscillators, respectively. The green broken line shows the simulation curve when the sum of intensities of the two Lorentz oscillators, $I$, is reduced to $0.5I$ $(-\Delta I(0.25\text{ps})/I = 0.5)$. **b,** $\varepsilon_2$ spectra. The black solid and purple broken lines show the experimental result and curve fitted with two Lorentz oscillators, respectively. The green broken line shows the simulation curve for $0.5I$. **c,** The blue and yellow solid lines indicate the two Lorentz oscillator components $I_{\text{LO1}}$ and $I_{\text{LO2}}$ used for fitting curves of the steady state spectra shown by the purple broken lines in (**a, b**). The blue and yellow broken lines indicate the components of $0.5I_{\text{LO1}}$ and $0.5I_{\text{LO2}}$, respectively, in the simulation curves of the transient spectra marked by green broken lines in (**a, b**). **d,** The estimated dependence of $-\Delta I(0.25\text{ps})/I$ and $n_c$ on $-\Delta R(0.25\text{ps})/R$. **e, f,** Schematics of a half-filled 1D Mott insulator state, (**e**) the ground state, and (**f**) an excited state with a doublon and holon.

**Table S2 | Fitting parameters of $R$ and $\varepsilon_2$ spectra for $Sr_2CuO_3$.**

| | $\hbar^2 S_{1,2}$ (eV$^2$) | $\hbar\omega_{1,2}$ (eV) | $\hbar\gamma_{1,2}$ (eV) | $\varepsilon_\infty$ |
|---|---|---|---|---|
| LO1 | 6.5 ± 0.1 | 1.76 ± 0.005 | 0.33 ± 0.005 | 2.7 ± 0.05 |
| LO2 | 4.3 ± 0.1 | 2.4 ± 0.05 | 1.2 ± 0.03 | |

Since it is difficult to dope carriers to $Sr_2CuO_3$ by element substitution, the same characteristic has not been experimentally demonstrated. However, it is natural to assume a similar trend to that observed in 2D cuprates. Namely, we can reasonably consider that both the spectral weight of absorption by excitons and that by free D-H pairs are transferred to the lower energies, and the transferred weight corresponds to the carrier number. Taking these facts into account, in the analysis, we assumed that the transition intensities of the two Lorentz components LO1 and LO2, $I_{\mathrm{LO1}}$ and $I_{\mathrm{LO2}}$, in equation (S4) decrease at the same rate when Ds and Hs are produced. Namely, $\Delta I_{\mathrm{LO1}}/I_{\mathrm{LO1}} = \Delta I_{\mathrm{LO2}}/I_{\mathrm{LO2}} = \Delta I/I$, where $I = I_{\mathrm{LO1}} + I_{\mathrm{LO2}}$ is the total transition intensity, and $\Delta I_{\mathrm{LO1}}$, $\Delta I_{\mathrm{LO2}}$, and $\Delta I$ are the changes in $I_{\mathrm{LO1}}$, $I_{\mathrm{LO2}}$, and $I$, respectively. For example, when $-\Delta I/I$ is set to 0.5, i.e., $I$ is reduced to $0.5I(= 0.5I_{\mathrm{LO1}} + 0.5I_{\mathrm{LO2}})$, $R$ and $\varepsilon_2$ are reduced, the spectra of which are shown by the green broken lines in Fig. S5(a) and (b), respectively. The $\varepsilon_2$ spectra of the two Lorentz components, LO1 and LO2, are shown by the blue and yellow broken lines, respectively, in Fig. S5(c). The relationship between $-\Delta R(0.25\mathrm{ps})/R$ and $-\Delta I/I$ is analysed by changing $-\Delta I/I$, as shown in Fig. S5(d).

Finally, we must convert the value of $-\Delta I/I$ to $n_{\mathrm{c}}$. In a simple 1D half-filled Hubbard system consisting only of $U$ and $t$, it has been theoretically shown that carrier doping reduces the spectral intensity of the absorption due to D or H, transitioning it to a Drude response irrespective of carrier density[18,19]. This can be explained by the concept of spin-charge separation characteristic of 1D system with large $U$[20]. This feature was also demonstrated experimentally by photoexciting the half-filled 1D Mott insulator ET-$F_2$TCNQ to generate carriers[21].

Figures S5(e) and (f) show schematics of the electron configurations in the ground state and excited states with one D and one H, respectively, when the system has large $U$ and carriers are strongly localized. It can be understood from these figures that introducing one D or one H reduces the absorption intensity by two sites. That is, the reduction in spectral intensity $I$ in the Mott gap transition region, $-\Delta I/I$, should be equal to $2n_{\mathrm{c}}$. For the case with finite $t$, as in actual systems, the theoretical calculation of the relation between $-\Delta I/I$ and the hole density $n_{\mathrm{h}}$ was performed for $U/t = 10$, which is a typical parameter for treating cuprates like $Sr_2CuO_3$[14]. The results showed the relation $-\Delta I/I \propto 2.6n_{\mathrm{h}}$, when the doping density is low. The situation should be identical for the electron doping (Fig. S5(f)). Therefore, the decrease of the spectral intensity, $-\Delta I/I$, can be approximated as $2.6n_{\mathrm{c}}$.

In our study, we assumed the simple relationship $-\Delta I/I = 2n_{\mathrm{c}}$ and converted the relation between $-\Delta R(0.25\mathrm{ps})/R$ and $-\Delta I/I$ (the left vertical axis) into a relation between $-\Delta R(0.25\mathrm{ps})/R$ and $n_{\mathrm{c}}$ using the right vertical axis in Fig. S5(d). From the latter relationship, we obtained the $|E_{\mathrm{MIR}}|$ dependence of $n_{\mathrm{c}}$ shown in Fig. 4g in the main text.

## S5 Propagation effects on HHs

In this section, we discuss the possibility of phase shifts arising from the propagation of the MIR electric field and the HHs. In this study, the HHs were measured in a reflection geometry. Consequently, the HH radiation emitted into free space is the coherent sum of HHs generated at different depths beneath the crystal surface and emitted at different times. HHs generated deeper inside the crystal are attenuated by absorption as they propagate toward the surface. During this propagation, the emitted HHs are also affected by changes in the optical constants at their respective photon energies induced by the MIR-driven modification of the electronic state. As a result, the propagation process may give rise to an MIR-field-dependent phase shift in the emitted HHs.

To estimate the magnitude of the phase shift of the HHs arising from changes in the optical constants associated with the photoinduced modification of the electronic state, we performed simulations using the geometry illustrated in Fig. S6(a). As shown in Fig. S6(a), the propagation directions of the MIR pulse and the HHs are antiparallel and parallel to the $z$ axis, respectively, and the crystal surface is defined as $z = 0$. The MIR electric field is assumed to propagate through the crystal without attenuation. Let $\xi$ denote the distance from the crystal surface to the position where an HH is generated, so that the generation point is located at $z = -\xi$. We further assume that, at time $t_0$, the electric field of an HH with angular frequency $\omega$ generated at the crystal surface ($\xi = 0$) is expressed at the surface ($z = 0$) as follows:

$$E_{\mathrm{HHG}}(\xi = 0, t) = E_0 \exp(-i\omega t) \tag{S5}$$

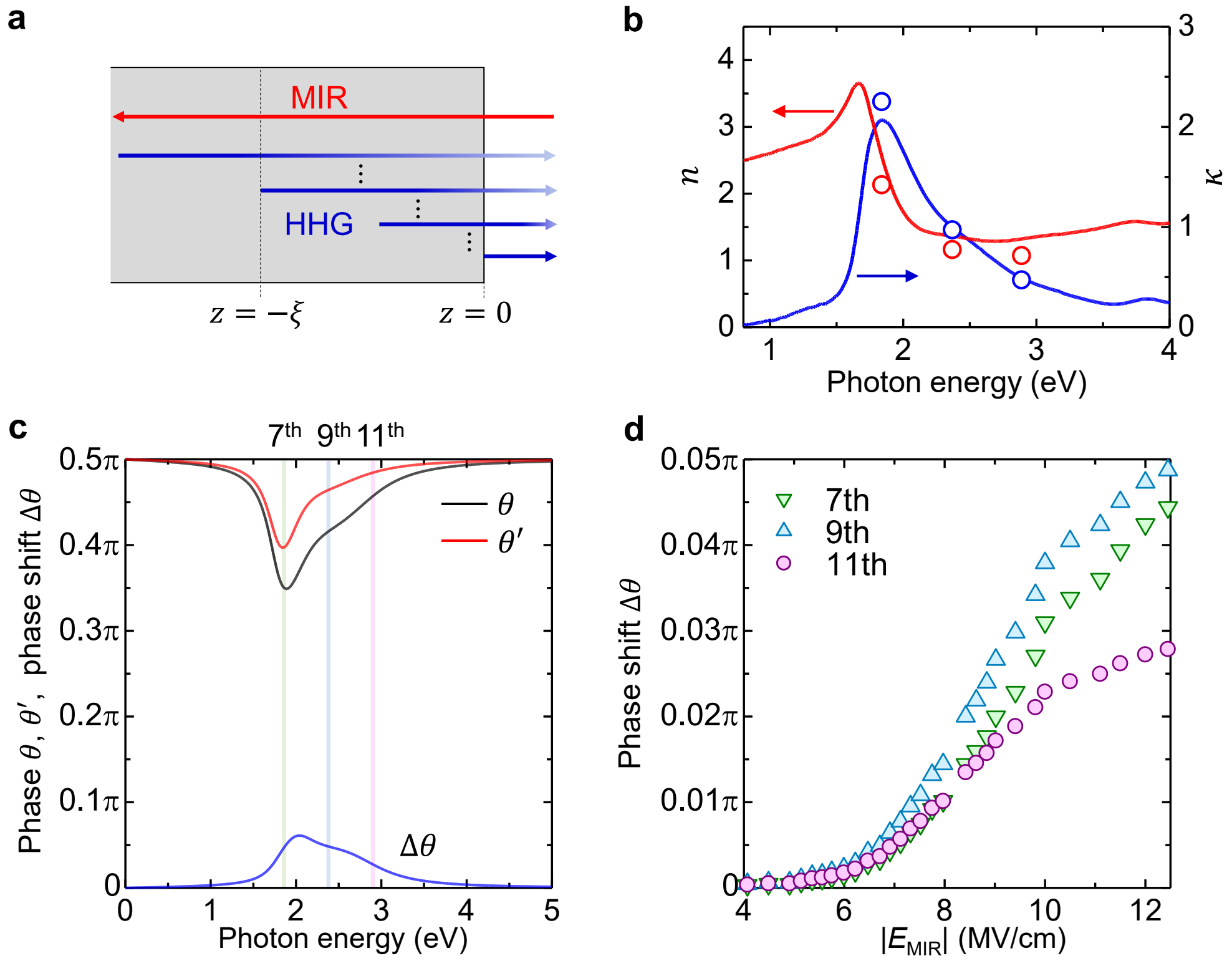

**Fig. S6 | Propagation effects on HHs. a**, Schematic illustration of HHG in the reflection geometry. **b**, Refractive index $n$ (red line) and extinction coefficient $\kappa$ (blue line) obtained from the steady-state reflectivity spectrum for light polarized along the $b$ axis. The red and blue circles represent the values of $n$ and $\kappa$ listed in Table S3 for the 7th, 9th, and 11th harmonic frequencies which are obtained by fitting the reflectivity spectrum with a two-Lorentz-oscillator model (Section S4). **c**, Phases of the HH fields at the crystal surface. $\theta$ is calculated from $n$ and $\kappa$ in Table S3, whereas $\theta'$ is calculated from $n'$ and $\kappa'$, where $n'$ and $\kappa'$ are obtained from the two-Lorentz-oscillator model with reduced oscillator strengths due to MIR-field-induced carrier generation (Section S4). $\Delta\theta$ denotes the transient phase shift caused by the change from ($n$, $\kappa$) to ($n'$, $\kappa'$) listed in Table S3. **d**, $|E_{\mathrm{MIR}}|$ dependence of $\Delta\theta$ for the 7$^{\mathrm{th}}$, 9$^{\mathrm{th}}$ and 11$^{\mathrm{th}}$ harmonics.

Next, as illustrated in Fig. S6(a), we consider an HH generated at $z = -\xi$. Assuming that the electric field of the HH reaches the crystal surface ($z = 0$) at time $t_0$ and taking the above initial assumption (S5) into account, the time dependence of the HH electric field at the surface is expressed as follows:

$$E_{\mathrm{HHG}}(\xi,t) = E_0 \exp\left[-i\omega\left(t - \frac{n_{\mathrm{MIR}}\xi}{c} - \frac{n}{c}\xi\right) - \frac{\omega\kappa}{c}\xi\right]$$

$$= E_0 \exp(-i\omega t)\cdot \exp\left\{[i(n_{\mathrm{MIR}} + n) - \kappa]\frac{\omega}{c}\xi\right\} \tag{S6}$$

Here, $n_{\mathrm{MIR}}$ is the refractive index at the MIR frequency, and $n$ and $\kappa$ are the refractive index and extinction coefficient of the HH at the corresponding harmonic frequency in the steady state, respectively. Since the HH electric field at the crystal surface is the coherent sum of the HHs generated at all positions in the crystal, it is given by the following integral:

$$E_{\mathrm{total\ HHG}} = \int_0^\infty E_{\mathrm{HHG}}(\xi,t)d\xi$$

$$= E_0 \exp(-i\omega t)\left[\frac{c}{[i(n+n_{\mathrm{MIR}}) - \kappa]\omega}\exp\left\{[i(n_{\mathrm{MIR}} + n) - \kappa]\frac{\omega}{c}\xi\right\}\right]_0^\infty$$

$$= E_0 \exp(-i\omega t)\frac{c}{[\kappa - i(n+n_{\mathrm{MIR}})]\omega}$$

$$= E_0 \exp(-i\omega t)\frac{c}{\omega}\cdot\frac{\kappa + i(n+n_{\mathrm{MIR}})}{\kappa^2 + (n+n_{\mathrm{MIR}})^2} \tag{S7}$$

Since the phase of an HH generated at the crystal surface at time $t_0$ is given by $-\omega t$, the additional phase $\theta$ arising from the coherent summation of HHs generated at different depths is expressed as

$$\theta = \tan^{-1}\left(\frac{n+n_{\mathrm{MIR}}}{\kappa}\right). \tag{S8}$$

To evaluate the phase shift caused by the transient modification of the electronic state due to photocarrier generation, we calculate the phase $\theta'$ after $n$ and $\kappa$ are modified by the photocarriers, and then evaluate the change in phase, $\Delta\theta = \theta' - \theta$. Denoting the refractive index and extinction coefficient after photocarrier generation by $n'$ and $\kappa'$, respectively, the phase shift is given by

$$\Delta\theta = \theta' - \theta = \tan^{-1}\left(\frac{n'+n_{\mathrm{MIR}}}{\kappa'}\right) - \tan^{-1}\left(\frac{n+n_{\mathrm{MIR}}}{\kappa}\right) \quad \text{(S9).}$$

In the present study, we did not measure the broadband transient reflectivity ($R$) spectrum under MIR excitation. Therefore, the refractive index $n'$ and extinction coefficient $\kappa'$ cannot be determined directly from the transient $R$ spectrum. Instead, we estimate these quantities using the results of the two-Lorentz-oscillator analysis. First, the values of $n$ and $\kappa$ at the photon energies corresponding to the 7th, 9th, and 11th harmonics were calculated from the parameters of the two Lorentz oscillators introduced to reproduce the steady-state $R$ spectrum (Table S3). These values are in good agreement with those obtained from the Kramers–Kronig analysis of the steady-state $R$ spectrum as shown in Fig. S6(b). The values of $n'$ and $\kappa'$ were then calculated from the parameters of the two Lorentz oscillators whose oscillator strengths were adjusted to reproduce the peak value of the $R$ spectrum under MIR excitation with different electric-field amplitudes. The details of this analysis are given in the main text and Section S4. Table S3 lists the values of $n'$ and $\kappa'$ at the photon energies corresponding to the 7th, 9th, and 11th harmonics for the maximum MIR electric-field amplitude, $|E_{\mathrm{MIR}}| = 12.5\ \mathrm{MV/cm}$. For $n_{\mathrm{MIR}}$, we used the corresponding value of 2.37 obtained from this analysis. Figure S6(c) shows the harmonic-energy dependence of $\theta$ (black line) and $\theta'$ (red line) for $|E_{\mathrm{MIR}}| = 12.5\ \mathrm{MV/cm}$, together with the corresponding phase shift $\Delta\theta$ (blue line). Finally, Fig. S6(d) shows the MIR electric-field dependence of $\Delta\theta$ for the 7th, 9th, and 11th harmonics, estimated from the electric-field dependence of $n'$ and $\kappa'$ obtained from the same two-Lorentz-oscillator analysis.

**Table S3 Refractive indices ($n$ and $n'$) and extinction coefficients ($\kappa$ and $\kappa'$) at the photon energies of HH lights.** $n$ and $\kappa$ are the steady-state values, while $n'$ and $\kappa'$ are the values under the MIR excitation with $|E_{\mathrm{MIR}}| = 12.5\ \mathrm{MV/cm}$.

| 1.84 eV (7th HHG) | | 2.37 eV (9th HHG) | | 2.89 eV (11th HHG) | |
|---|---|---|---|---|---|
| Steady state | MIR exc. | Steady state | MIR exc. | Steady state | MIR exc. |
| $n$ =2.13 | $n'$ =1.73 | $n$ =1.16 | $n'$ =1.32 | $n$ =1.07 | $n'$ =1.36 |
| $\kappa$ =2.25 | $\kappa'$ =1.38 | $\kappa$ =0.97 | $\kappa'$ =0.42 | $\kappa$ =0.47 | $\kappa'$ =0.18 |

As shown in Fig. S6(c, d), the phase shift of the HHs induced by MIR excitation is at most $0.05\pi$ for all harmonic orders. This value is much smaller than the phase shift predicted by the DMFT calculations, which reaches nearly $\pi$ (Fig. 5k, l). In the above analysis, we neglected the attenuation of the MIR pulse as it propagates into the crystal. In reality, photocarrier generation is expected to enhance the attenuation of the MIR pulse, thereby reducing its penetration depth. Consequently, the contribution of HHs generated at larger values of $\xi$ is expected to be smaller than that estimated here. The actual phase shift $\Delta\theta$ should therefore be even smaller than that shown in Fig. S6(c). These results indicate that the phase shift of the HHs arising from changes in the optical constants induced by the photoinduced modification of the electronic state is small compared with that obtained by the theoretical simulations in Fig. 5k, l. Furthermore, this phase shift originates from photocarrier generation, which develops gradually over several cycles of the MIR field, as shown in Fig. 5e, f. Such a small and gradual phase evolution cannot account for the large redshift experimentally observed for the HHs. We therefore conclude that the observed redshift of the HHs does not originate from propagation effects associated with photoinduced changes in the optical constants.

## S6 Supplementary data of DMFT simulations

This section presents supplementary data for the DMFT simulation of the single-band Hubbard model. The parameter sets for the system and pump pulse are the same as those mentioned in the main text, unless specified otherwise.

### S6.1 Details of the HHG spectra

Figure S7 shows the HH spectra calculated using DMFT as a function of the field amplitude $|E_{\mathrm{MIR}}|$, which correspond to the experimental HH spectra shown in Fig. 2b and 4a–f in the main text. With increase in $|E_{\mathrm{MIR}}|$, the peak frequencies of the calculated $n$th harmonic spectra around $\omega = n\Omega$, $\omega_{\mathrm{peak},n}$, exhibit redshifts for $n > 5$, while they remain nearly unchanged for $n = 3$, consistent with the experimental observations. The behaviour of the peak for $n = 5$ represents an intermediate case.

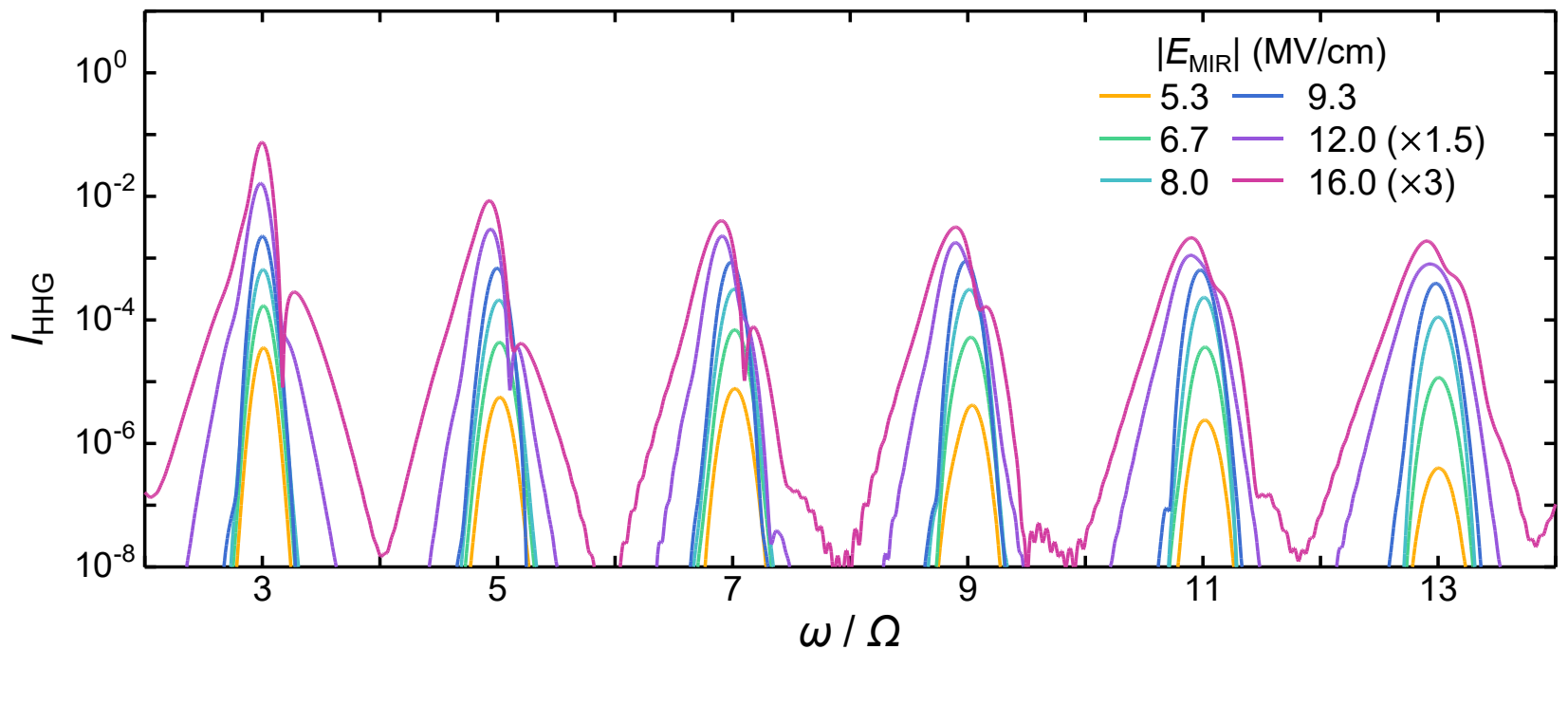

**Fig. S7 | HH spectra calculated by DMFT.** The parameter values of the model and pump pulse are described in the Method section in the main text.

Figure S8(a) shows the $|E_{\mathrm{MIR}}|$ dependence of the intensity of each harmonic, $I_{n\mathrm{HG}}$, which is evaluated by integrating $I_{\mathrm{HHG}}(\omega)$ around $\omega = n\Omega$. The overall trend is consistent with the experimental results. Specifically, for $n = 3$, the $I_{3\mathrm{HG}}$ deviates upward from the scaling expected in the perturbation theory, $I_{3\mathrm{HG}} \propto |E_{\mathrm{MIR}}|^6$ (Fig. S8(b)). This enhancement can be attributed to a reduction in the Mott gap and an increase in the carrier density. The Mott-gap reduction enhances in-gap off-resonance processes, which dominate the third-harmonic generation in the perturbative regime, while the increase in carrier density enhances the contribution of intraband currents to the in-gap radiation. For $n = 5$, the intensity follows the perturbation theory as $I_{5\mathrm{HG}} \propto |E_{\mathrm{MIR}}|^{10}$ at low electric fields and then saturates. In contrast, for $n \geq 7$, the deviation from the perturbative scaling is downward, which appears even at low electric-field amplitudes (Fig. S8(b)). These trends are consistent with the experimental findings (Fig. 2c).

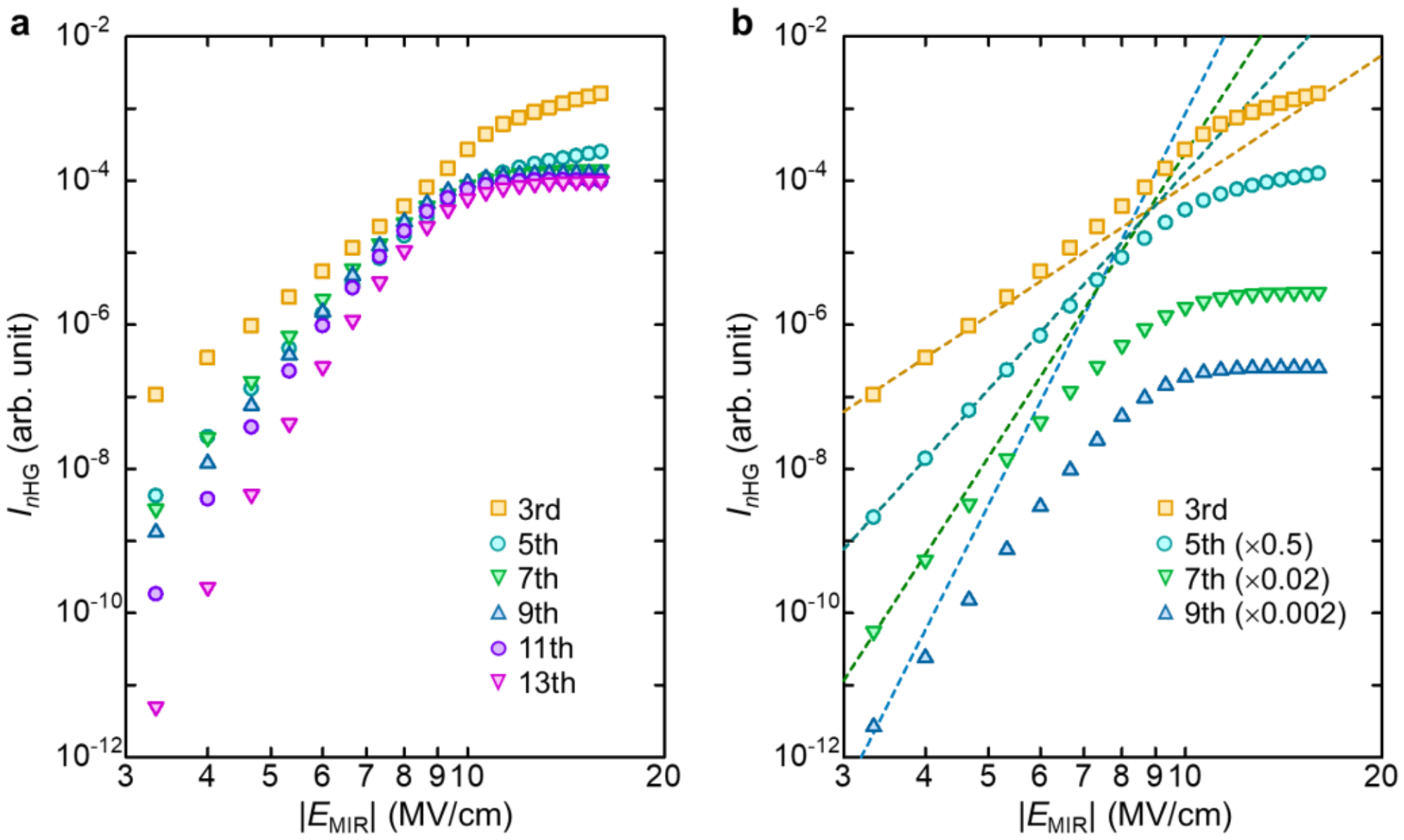

**Fig. S8 | $|\boldsymbol{E}_{\mathbf{MIR}}|$ dependence of the integrated intensities of HHs calculated by DMFT. a,** Full data from the 3rd to 13th harmonics. **b,** Data from the 3rd to 9th harmonics multiplied by the factors shown in the panel. The broken lines show the relations expected from the perturbation theory ($I_{n\mathrm{HG}} \propto |E_{\mathrm{MIR}}|^{2n}$).

## S6.2 Supplementary analyses for sub-cycle radiation

To extract the phases of sub-cycle radiations on a cycle-by-cycle basis, we applied a temporal window of half-cycle width $T_{\mathrm{hp}} = \pi/\Omega$ to the current. Specifically, we used the following window function:

$$F_{\mathrm{box}}(x,\eta)=\frac{1+\tanh\left(-\frac{2x-1}{2\eta}\right)\tanh\left(\frac{2x+1}{2\eta}\right)}{2}, \tag{S10}$$

which smoothly covers the range $x\in[-0.5,0.5]$ with a smoothing width of $\eta$. To focus on the radiation around $t=t_{\mathrm{p}}$, we defined the sub-cycle differentiated current as follows:

$$\dot{J}(t,t_{\mathrm{p}})=F_{\mathrm{box}}\left(\frac{t-t_{\mathrm{p}}}{T_{\mathrm{hp}}},\eta\right)\dot{J}(t), \tag{S11}$$

which extracts the current around $t\in\left[t_{\mathrm{p}}-\frac{T_{\mathrm{hp}}}{2},t_{\mathrm{p}}+\frac{T_{\mathrm{hp}}}{2}\right]$. When $\eta$ is sufficiently small, the total current can be expressed as $\dot{J}(t)=\sum_l\dot{J}(t,t_{\mathrm{ref}}+lT_{\mathrm{hp}})$, and similarly in the frequency domain as $\dot{J}(\omega)=\sum_l\dot{J}(\omega,t_{\mathrm{ref}}+lT_{\mathrm{hp}})$, where $t_{\mathrm{ref}}$ is a constant that adjusts the central position within each time window. Constructive or destructive interference between the different sub-cycle radiations $\dot{J}(\omega,t_{\mathrm{ref}}+lT_{\mathrm{hp}})$ leads to peaks or valleys in $|\dot{J}(\omega)|$ and in the corresponding HH spectrum $I_{\mathrm{HHG}}(\omega)=|\dot{J}(\omega)|^2=|\omega J(\omega)|^2$. We set $T_{\mathrm{hp}}=7.6\mathrm{fs}(\hbar\Omega=0.263\mathrm{eV})$, $\eta=0.2$, and $t_{\mathrm{ref}}=-1.59$ fs. In the range of $t\sim-90$ to $90$ fs, 21 windows($l=-10$ to $10$) exist, as indicated by the coloured lines in Fig. S9, in which only windows with even numbers are shown. For these parameters, $\sum_l F_{\mathrm{box}}\left(\frac{t-t_{\mathrm{ref}}-lT_{\mathrm{hp}}}{T_{\mathrm{hp}}},\eta\right)\cong 1$, as shown by the grey solid line in the same figure.

In Fig. S10, the difference of $\mathrm{Arg}[\dot{J}(\omega,t_{\mathrm{ref}}+lT_{\mathrm{hp}})]$ from $\mathrm{Arg}[\dot{J}(\omega,t_{\mathrm{ref}}+lT_{\mathrm{hp}})]$ at $l=-9$, $\Delta\mathrm{Arg}[\dot{J}(\omega,t)]$, at $\omega=n\Omega$ and $\omega=\omega_{\mathrm{peak},n}$ $(n=3-11)$ for each half cycle are plotted. The $|E_{\mathrm{MIR}}|$ values are 6.7, 12.0, and 16.0 MV/cm. The marker size indicates the intensity of $|\dot{J}(\omega,t_{\mathrm{ref}}+lT_{\mathrm{hp}})|$. The data for the 3rd and 9th harmonics are shown in Figs. 5j–l in the main text. At $E_{\mathrm{MIR}}=6.7$ MV/cm, the phase evolutions for $\omega=n\Omega$ and $\omega=\omega_{\mathrm{peak},n}$ are similar, and only slight differences appear in each cycle (see panels (a–e)). In contrast, at $E_{\mathrm{MIR}}=12.0$ and 16.0 MV/cm, the phase difference remains small for $n=3$, which corresponds to the in-gap radiation (see panels (f) and (k)). However, for higher harmonics ($n\geq 5$), a significant deviation arises between $\omega=n\Omega$ and $\omega=\omega_{\mathrm{peak},n}$; at $\omega=n\Omega$, finite phase variations occur depending on $t$, while at $\omega=\omega_{\mathrm{peak},n}$, phase variations are relatively small, particularly when the corresponding intensity is strong. This indicates that the radiation at $\omega = \omega_{\mathrm{peak},n}$ accumulates more constructively over cycles than that at the formal harmonic frequency $\omega = n\Omega$.

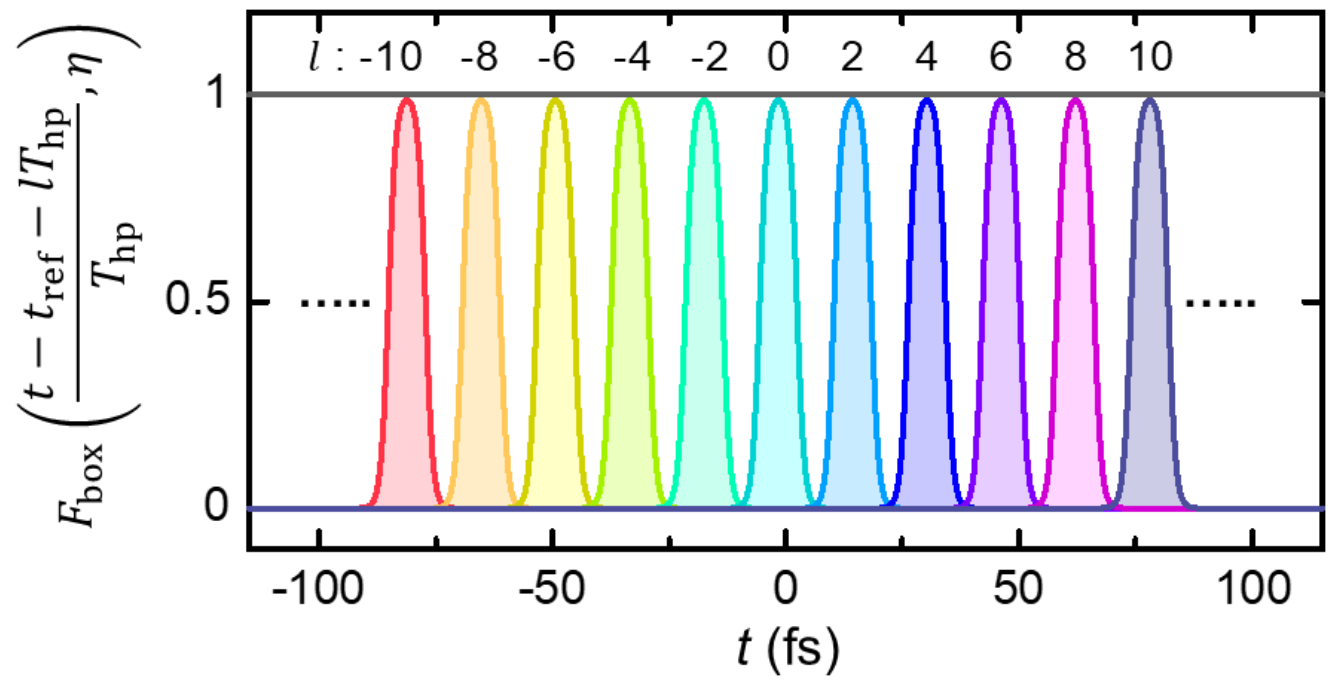

**Fig. S9 | Window functions $F_{\mathrm{box}}\left(\frac{t-t_{\mathrm{ref}}-lT_{\mathrm{hp}}}{T_{\mathrm{hp}}},\eta\right)$ for different values of $l$.** Here, $t_{\mathrm{ref}} = -1.59$ fs and $\eta = 0.2$. The grey line shows $F_{\mathrm{tot}} \equiv \sum_l F_{\mathrm{box}}\left(\frac{t-t_{\mathrm{ref}}-lT_{\mathrm{hp}}}{T_{\mathrm{hp}}},\eta\right)$, which is practically 1.

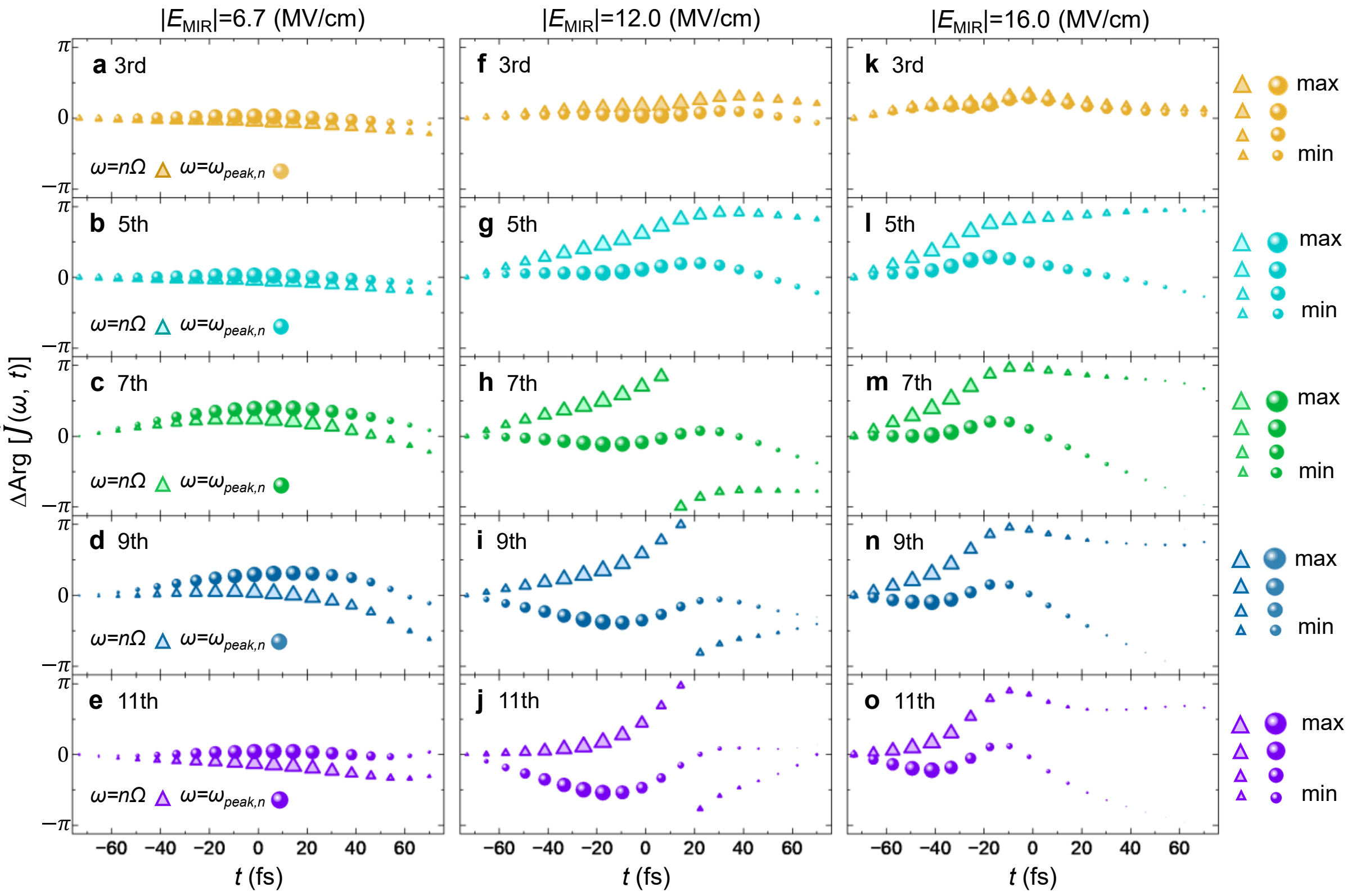

**Fig. S10 | Phase changes for the 3rd to 11th harmonics, $\Delta\mathrm{Arg}[\dot{J}(\boldsymbol{\omega},\boldsymbol{t})]$.** The size of a marker is proportional to $|\dot{J}(\omega,t)|$. **(a–e)** $|E_{\mathrm{MIR}}| = 6.7$ MV/cm, **(f–j)** $|E_{\mathrm{MIR}}| = 12.0$ MV/cm, and **(k–o)** $|E_{\mathrm{MIR}}| = 16.0$ MV/cm.

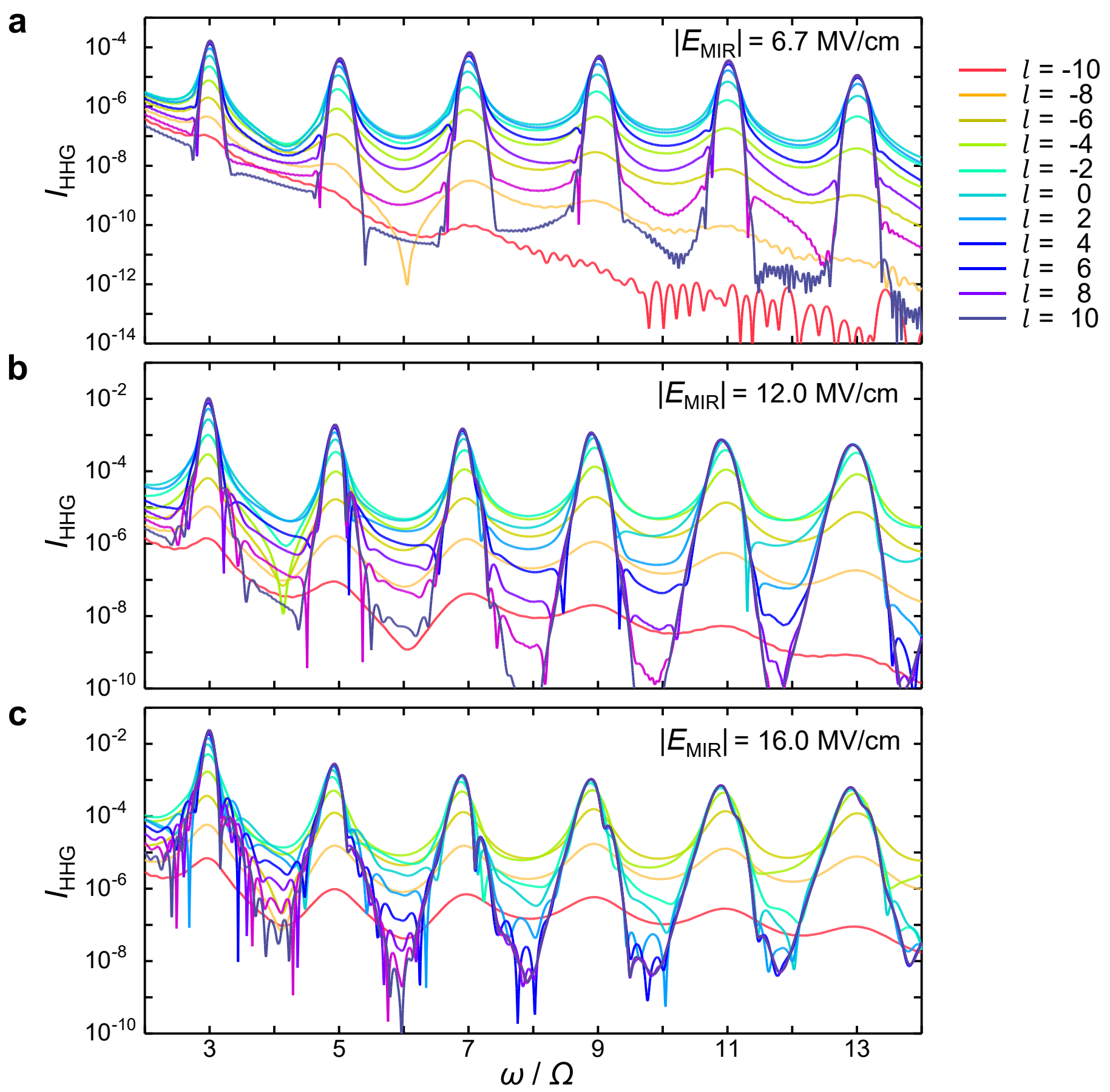

**Fig. S11 Truncated radiation spectra $\tilde{I}_l(\omega, t_{\mathrm{ref}})$ for different values of $l$.** (**a**) $|E_{\mathrm{MIR}}| = 6.7\ \mathrm{MV/cm}$, (**b**) $|E_{\mathrm{MIR}}| = 12.0\ \mathrm{MV/cm}$, and (**c**) $|E_{\mathrm{MIR}}| = 16.0\ \mathrm{MV/cm}$. Each spectrum shows the radiation accumulated up to $t_{\mathrm{ref}} + lT_{\mathrm{hp}}$.

To investigate the cycle-by-cycle accumulation of HH intensity, the following truncated current was introduced:

$$\tilde{j}_l(t, t_{\mathrm{ref}}) = \sum_{k \le l} j\big(t, t_{\mathrm{ref}} + iT_{\mathrm{hp}}\big), \tag{S12}$$

which corresponds to the current accumulated up to $t = t_{\mathrm{ref}} + lT_{\mathrm{hp}}$. The corresponding radiation spectrum is given by $\tilde{I}_l(\omega, t_{\mathrm{ref}}) = \left|\tilde{j}_l(\omega, t_{\mathrm{ref}})\right|^2$. The results for even $l$ are shown in Fig. S11. This figure conceptually corresponds to the schematics shown in Fig. 1a in the main text. For small values of $l$, the peak structures in the spectrum are not well developed; clear peak structures emerge with increasing $l$. Peaks appear at $\omega = n\Omega$ ($n$:

odd number) for $E_{\mathrm{MIR}} =$ 6.7 MV/cm, whereas they develop at the lower frequencies ($\omega < n\Omega$) for $E_{\mathrm{MIR}} =$ 12.0 and 16.0 MV/cm. These observations are consistent with the above-mentioned discussion on the radiation phase shown in Fig. S10.

### S6.3 Carrier envelope phase (CEP) dependence of the HH spectrum and antiferromagnetic (AFM) order

Figure S12 shows the dependence of the HH spectrum and AFM order $m_{\mathrm{AFM}}$ on the CEP of the MIR pump pulse. Here, the CEP values of the MIR electric field, $\phi_{\mathrm{CEP}}$, are varied in steps of $\pi/8$ over the range of $[-\pi, \pi]$. Neither quantity depends upon the CEP values. Therefore, unlike other HH shift phenomena previously observed in semiconductors and topological insulators[22,23], the HH peak redshifts discussed in this study are not sensitive to CEP.

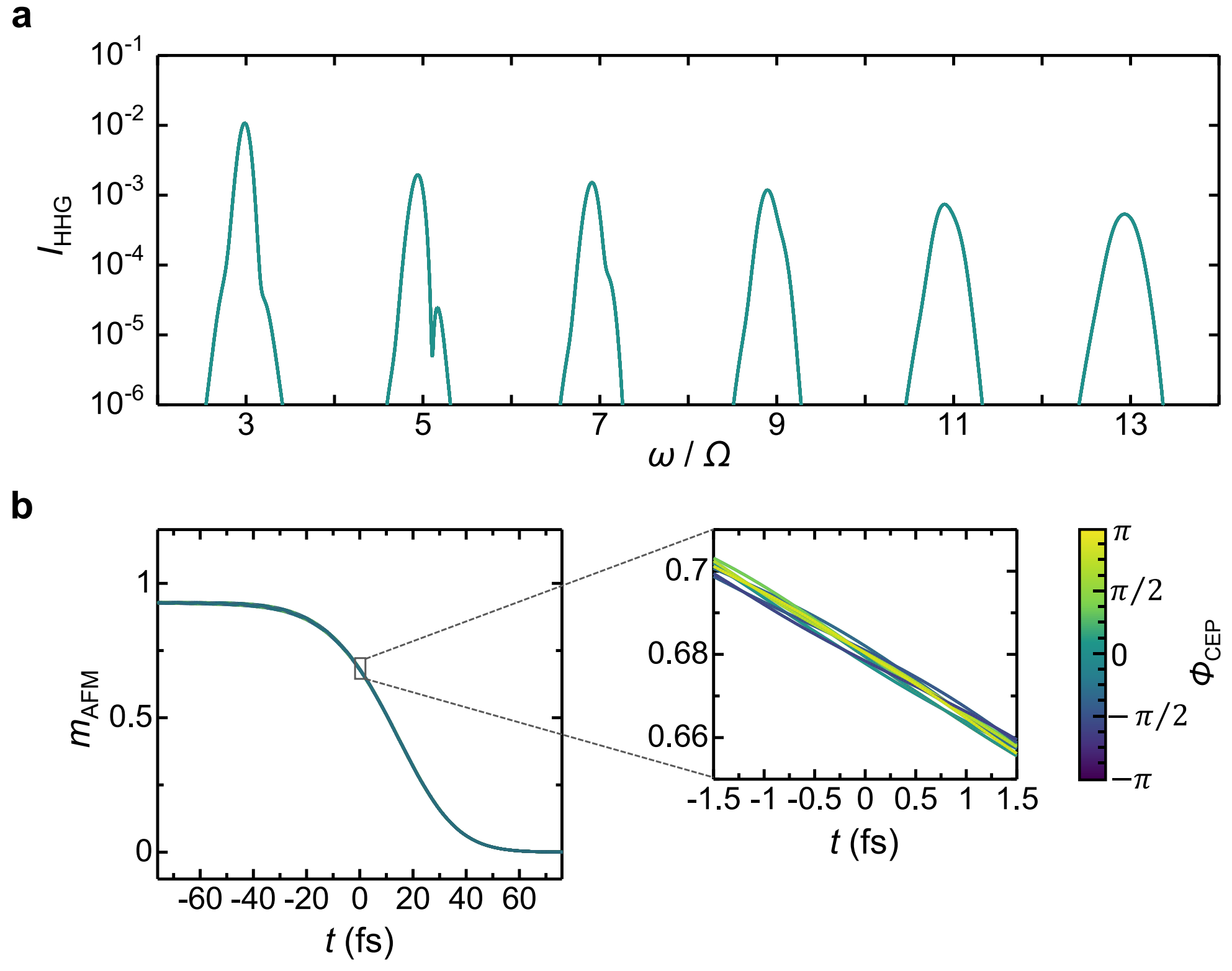

**Fig. S12 | HH spectra and time dependence of the AFM order $m_{\mathrm{AFM}}$ for different values of CEP, $\phi_{\mathrm{CEP}}$, simulated by DMFT.** $|E_{\mathrm{MIR}}| =$ 12.0 MV/cm, and the $\phi_{\mathrm{CEP}}$ is changed from $-\pi$ to $\pi$ in steps of $\pi/8$. The results in (**a**) and (**b**) almost completely overlap with each other. In (**b**), a magnified view of the left panel is provided in the right panel to highlight the differences.

### S6.4 Details of time-evolution of single-particle spectra

Here, we provide a more detailed description of the band-structure evolution shown in Fig. 5 of the main text. Figure S13 illustrates the time-dependent changes in the band structure under the applied electric field.

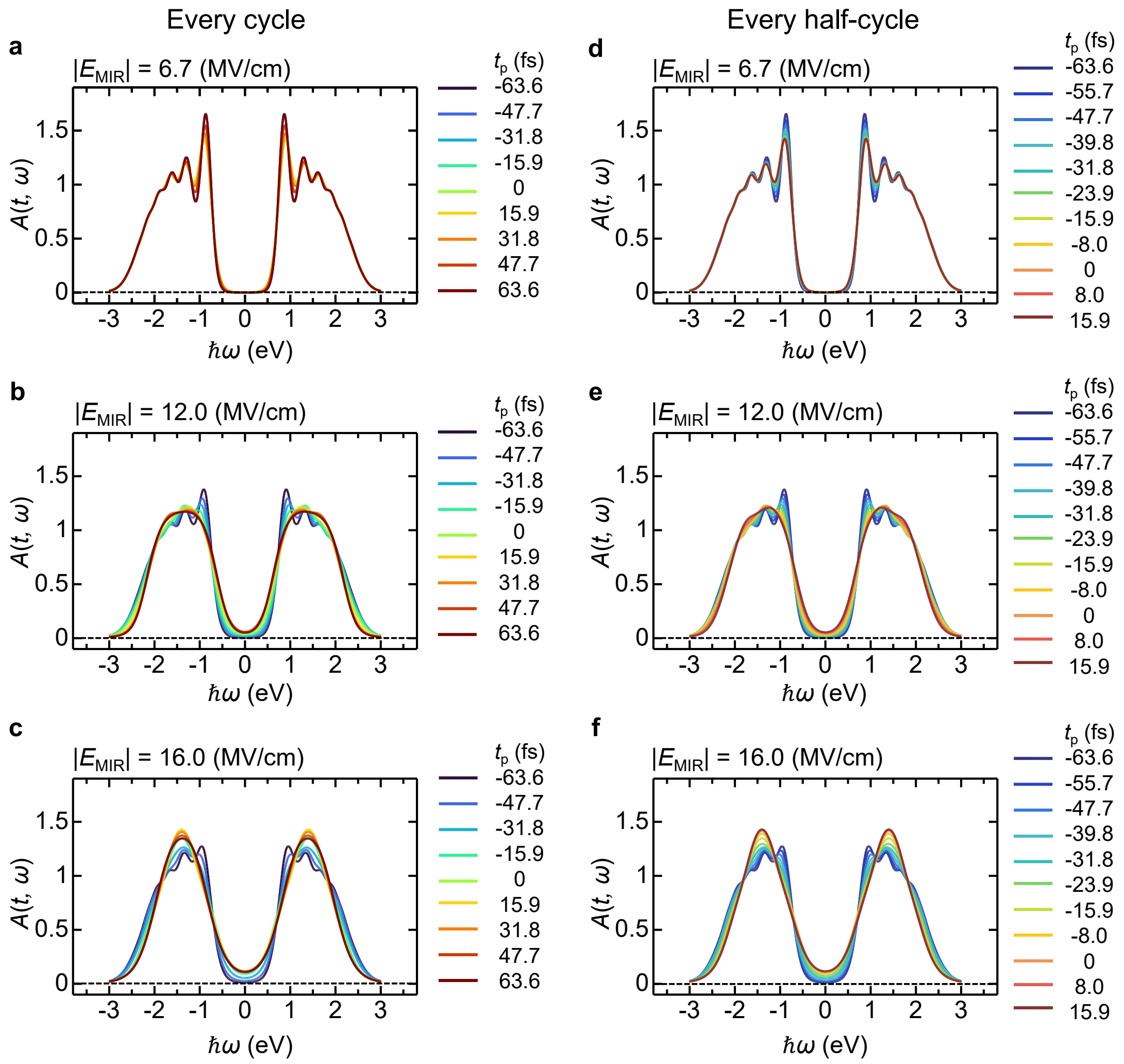

**Fig. S13 | Detailed views of the band-structure evolution $A(t_\mathrm{p}, \omega)$ presented in Fig. 5 of the main text.** The left panels (**a-c**) display the band structure for each optical cycle over the time interval from $t_p = t_0 - 4T_p$to $t_p = t_0 + 4T_p$. The right panels (**d-f**) show the band structure resolved at every half cycle, covering the time window from $t_p = t_0 - 4T_p$ to $t_p = t_0 + T_p$.

For the weak-field case $|E_\mathrm{MIR}| = 6.7$ MV/cm, the band shape remains nearly unchanged from cycle to cycle and returns to its original form once the pulse has passed.

In contrast, for stronger fields $|E_{\mathrm{MIR}}| = 12.0$ MV/cm and $|E_{\mathrm{MIR}}| = 16.0$ MV/cm substantial modifications of the band structure already appear in the temporal region preceding the pulse center $t < 0$, where the emission is strong. After the pulse, the Mott gap is reduced or even collapses, and the resulting band structure differs significantly from that before the excitation. Because the system is driven far from equilibrium, the equilibrium relations among carrier density, antiferromagnetic correlations, and band shape do not directly apply. Consequently, within the present nonequilibrium setup, the electronic structure can exhibit large changes even when the carrier number and AFM correlations vary only moderately. These observations support the interpretation that the cycle-to-cycle aperiodic evolution of the band structure accompanied by the melting of the Mott insulator is closely linked to the observed peak shifts.

## S7 Analysis with phenomenological doublon-holon model

In this section, we introduce a phenomenological model termed the doublon-holon model, that focuses on the dynamics of doublons and holons, and use it to clarify how the carrier distribution, band-structure renormalization, and relaxation-time variations influence HHG. In the DMFT analysis, these effects are incorporated simultaneously, which makes it difficult to disentangle their individual contributions. By contrast, the present doublon-holon model enables us to isolate these factors. As shown below, our analysis demonstrates that the observed HH peak shifts originate from band-structure modifications associated with the renormalization of the Mott gap.

The doublon-holon model is formulated as a one-dimensional two-band spinless fermion system, where each band corresponds to doublons or holons, i.e. elementary excitations emerging out of many-body correlations. The construction of the model is motivated by the fact that the three-step picture of HHG remains applicable even in a Mott insulator when the dynamics of elementary excitations is concerned. The Hamiltonian of this model is expressed as

$$\hat{H}(t) = \sum_k \left[\hat{c}_{k,0}^\dagger \; \hat{c}_{k,1}^\dagger\right] h_k(t) \begin{bmatrix} \hat{c}_{k,0} \\ \hat{c}_{k,1} \end{bmatrix}, \tag{S13}$$

where

$$h_k(t) = \begin{bmatrix} \epsilon_0(k - A(t), t) & -E(t)d_{01}\big(k - A(t)\big) \\ -E(t)d_{01}\big(k - A(t)\big) & \epsilon_1(k - A(t), t) \end{bmatrix}. \tag{S14}$$

Here, $k$ and $A(t)$ denote the momentum and the vector potential, respectively. The structure of this model closely parallels that of the conventional two-band semiconductor model; however, band 0 (1) corresponds to the doublon (holon). Their dispersions, $\epsilon_0(k)$ and $\epsilon_1(k)$, are identified with the doublon and holon dispersions $\epsilon_D(k)$ and $\epsilon_H(k)$ of the Hubbard model. In addition, we allow the dispersions to evolve in time to mimic the photoinduced melting of the Mott gap, as discussed below.

More specifically, the dispersions of holons and doublons in the one-dimensional Hubbard model are obtained from the Bethe ansatz[24]. A holon (doublon) is parametrized by a rapidity $\xi \in \mathbb{R}$, with the momentum $k_H(\xi)$ $(k_D(\xi))$ and energy $\epsilon_H(\xi)$ $(\epsilon_D(\xi))$ given by

$$k_H(\xi) = k_{\mathrm{D}}(\xi) + \pi = \frac{\pi}{2} - \xi - 2\int_0^{\infty} \frac{d\omega}{\omega} \frac{\mathcal{J}_0(\omega)\sin(\omega \sin\xi)}{1 + \exp(U\omega/2)}, \tag{S15}$$

$$\epsilon_{\mathrm{H}}(\xi) = \epsilon_{\mathrm{D}}(\xi) = \frac{U}{2} + 2\cos\xi + 2\int_0^{\infty} \frac{d\omega}{\omega} \frac{\mathcal{J}_1(\omega)\cos(\omega \sin\xi)\, e^{-U\omega/4}}{1 + \exp(U\omega/2)}. \tag{S16}$$

Here, $\mathcal{J}_n$ denotes the Bessel function of the first kind of order $n$. In equations (S15) and (S16), the hopping amplitude of the Hubbard model is set to $t_{\mathrm{hop}} = 1$ and $U$ denote the on-site Coulomb interaction.

To mimic the doublon–holon dynamics relevant for high-harmonic generation in the Hubbard model, we identify $\epsilon_0\left(k - \frac{\pi}{2}\right) = \epsilon_D(k)$ , and $\epsilon_1\left(k - \frac{\pi}{2}\right) = -\epsilon_D(-k)$ . Furthermore, $d_{01}(k)$ represents the effective dipole matrix element between the doublon and holon bands. To qualitatively reproduce the HHG spectra of the Hubbard model, we choose $d_{01}(k) = \alpha + \beta\cos k$, with $\alpha = 0.7$ $[e\text{Å}]$ and $\beta = 0.86$ $[e\text{Å}]$. We note, however, that the precise choice of $d_{01}(k)$ has little influence on the peak positions, which are the main focus of the present study.

The model is solved using the von Neumann equation for the single-particle density matrix, supplemented by phenomenological relaxation terms:

$$\rho_{ab,k}(t) = \langle \hat{c}_{kb}^{\dagger}\hat{c}_{ka}(t) \rangle, \tag{S17}$$

$$\partial_t \rho_k(t) = i[\rho_k(t), h_k(t)] - \frac{\rho_{\mathrm{diag},k}(t) - \rho_{\mathrm{eq},k}(t)}{T_1} - \frac{\rho_{\mathrm{off},k}(t)}{T_2(t)}. \tag{S18}$$

Correlation effects are incorporated phenomenologically through (i) a time-dependent renormalization of the band structure and (ii) time-dependent relaxation times. These assumptions are physically motivated, since photoexcitation modifies the carrier population and thereby the effective parameters governing elementary excitations. Accordingly, the temporal evolution of the band structure is modeled by introducing an effective time dependence of the interaction parameter $U(t)$ in the Bethe-ansatz dispersion equations (S15) and (S16). Although the population relaxation time $T_1$ should also vary, we find that its specific value and variation do not affect the physics discussed here.

Using this model, we compare cases with and without band-structure renormalization and relaxation-time variation, and analyze the effects of carrier redistribution, dispersion renormalization, and coherence on HHG. To match the bandwidth and gap obtained from the DMFT calculations, we set $t_{\mathrm{hop}} =$ 0.52 eV and choose an initial value $U = 3.12$ eV. The relaxation time $T_1$ is fixed to 1.3 ps.

Figures S14(a) and (b) show the carrier density and HHG spectra for different electric-field strengths in the absence of band-structure and relaxation-time renormalization. Although the generated carrier density $n_{\mathrm{c}}$ qualitatively agrees with the DMFT and experimental results, no shift of the HH peak is observed as the field strength increases. Since this situation corresponds to a rigid-band semiconductor model, these results indicate that a rigid-band picture cannot account for the experimentally observed HH peak shift.

Guided by the DMFT results, we next consider the case in which both the band gap and relaxation times decrease under strong driving and compare it with the unrenormalized case. Practically, we use a ramping function $F_{ramp}(t - t_{mod}, \sigma_{mod})$ to represent the time dependence of $T_2(t)$ and $U(t)$ as $T_2(t) = T_2(0) + \Delta T_2\, F_{ramp}(t - t_{mod}, \sigma_{mod})$ and $U(t) = U(0) + \Delta U\, F_{ramp}(t - t_{mod}, \sigma_{mod})$, where $t_{mod}$ is the center of the ramp and $\sigma_{mod}$ is the width of the ramp. $F_{ramp}(t, \sigma)$ takes the form of $\frac{1}{2} - \frac{3}{4}\sin\left(\pi\frac{t}{\sigma}\right) + \frac{1}{4}\sin^3\left(\pi\frac{t}{\sigma}\right)$ for $t \in \left[-\frac{\sigma}{2}, \frac{\sigma}{2}\right]$, and takes $0(1)$ for $t \leq -\frac{\sigma}{2}\left(t \geq \frac{\sigma}{2}\right)$. We choose $t_{mod} =$ 0.0 fs and $\sigma_{mod} =$ 63.3 fs, referring to the DMFT simulation. Figure S14(c) shows the corresponding evolution of the Mott gap $\Delta_\mathrm{Mott}$ and the coherence time $T_2$ for the three cases studied here. Figure S14(d) displays the corresponding HHG spectra for $E = 12.0$ MV/cm. We find that the HH peak exhibits a clear red shift when both the gap and relaxation times are reduced, consistent with the DMFT results. By contrast, a change in the relaxation times alone does not reproduce the red shift.

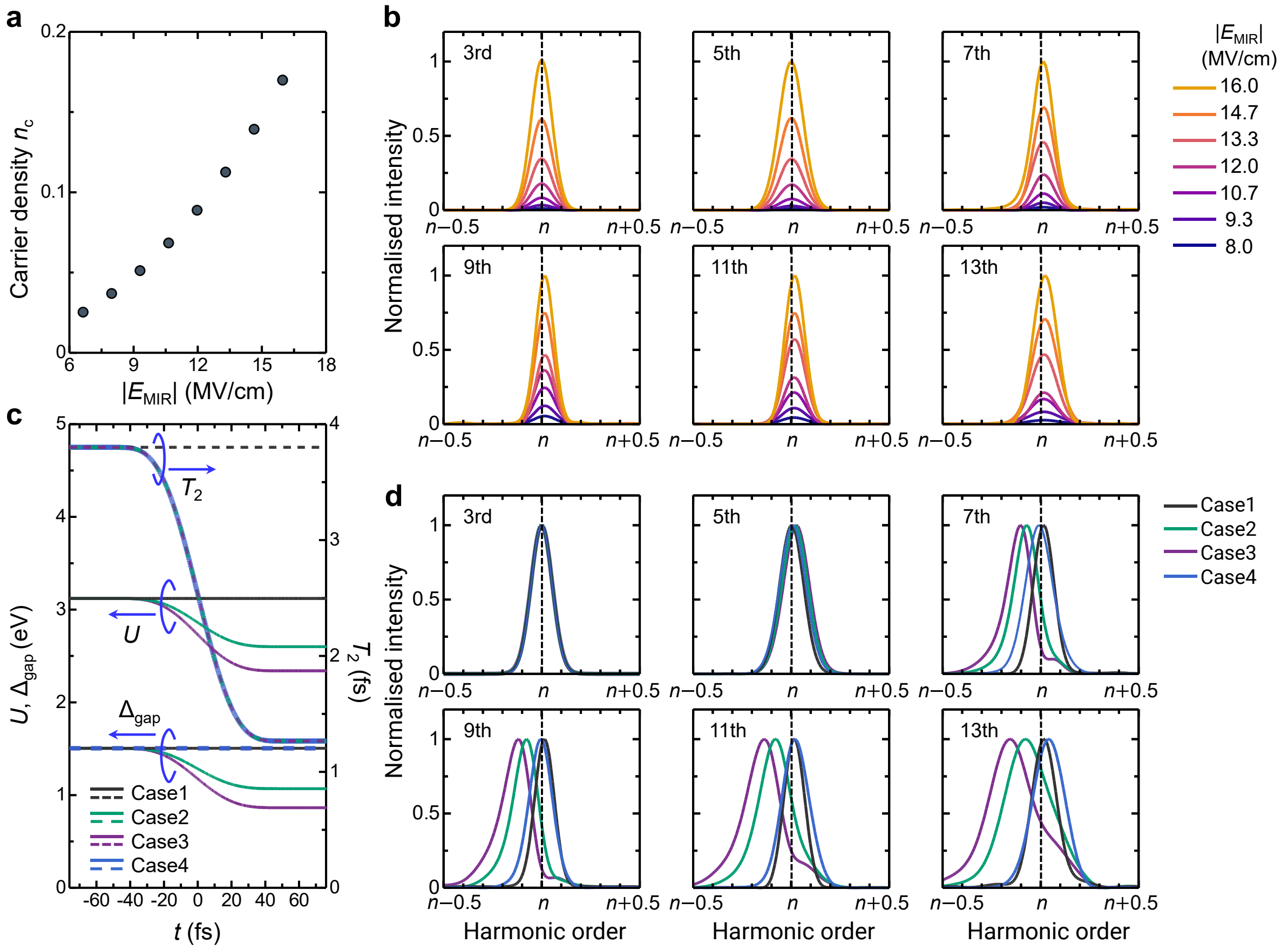

**Fig. S14 | Results of the analysis based on the DH model.** (**a**) Electric-field dependence of the carrier density $n_\mathrm{c}$ after the MIR pulse excitation in the absence of any parameter renormalization. (**b**) Electric-field dependence of the spectral shapes of high-harmonic peaks without parameter renormalization. (**c**) Time dependence of the Coulomb interaction energy $U$, the Mott gap energy $\Delta_\mathrm{Mott}$, and the relaxation time $T_2$ for the three cases compared in this study. In Case 1, the system parameters $U$ and $T_2$ are kept constant. In Cases 2 and 3, $U$ (the Mott gap) and $T_2$ are modulated in time, to be consistent with DMFT calculations. The modulation amplitude of $U$ is larger in Case 3 than in Case 2, resulting in correspondingly larger high-harmonic shifts shown in (d). In Case 4, only $T_2$ is modulated in time. (**d**) Spectral shapes of high-harmonic peaks corresponding to the cases shown in (c). Each spectrum is normalized by its maximum value. In all the calculations, the DH-model parameters are set to $t_\mathrm{hop}$ =0.52 eV and $U(0) = 3.12$eV. The MIR pulse is characterized by $t_0 = 0$ fs, $\sigma_0 =$ 63.3 fs, and $\hbar\Omega =$ 0.26 eV.

These results support our main conclusion from the DMFT analysis: the red shift of the HH peak originates from the dynamical modification of the band structure.